\documentclass[twoside,12pt]{article}
\usepackage{epsfig}

\def\Journal#1#2#3#4{{#1} {#2} (#4) #3 }

\def\PRO{{\em Prog. Theor. Phys.}}
\def\NPB{{\em Nucl. Phys.} B}

\def\PLB{{\em Phys. Lett.} B}

\def\PRL{\em Phys. Rev. Lett.}
\def\PREV{\em Phys. Rev.}
\def\PREP{\em Phys. Rep.}

\def\PRD{{\em Phys. Rev.} D}

\def\ANNP{\em Ann. Phys. (N.Y.)}
\def\RMP{{\em Rev. Mod. Phys.}}

\def\NATURE{{\em Nature}}
\def\NIMA{{\em Nucl. Instrum. Methods. Phys. Res.} Sect. A}
\def\EPJC{{\em Eur. Phys. J.} C}
\def\JHEP{{\em J. High Energy P.}}
\def\IJMPA{{\em Int. J. Mod. Phys.} A}
\def\JJAP{{\em Jpn. J. Appl. Phys.}}
\def\JPHYS{{\em J. Phys.} Conf. Ser.}
\def\JPHYSG{{\em J. Phys.} G}
\def\RVNC{{\em Riv. Nuovo Cim.}}
\def\APPB{{\em Acta Physica Polonica} B}
\def\ARNPS{{\em Ann. Rev. Nucl. Part. Sci.}}
\def\POS{{\em PoS}}
\def\PPCL{\em Proc. Phys. Soc. London Sect. A}
\def\PHYSA{{\em Physica} A}
\def\PAN{{\em Phys. Atom. Nucl.}}
\def\KOTO{K$^O$TO\ }
\def\DAFNE{DA$\Phi$NE\ }
\def\Kp{$K^+$\ }
\def\Km{$K^-$\ }
\def\Kpm{$K^{\pm}$\ }
\def\KL{$K^0_L$\ }
\def\KS{$K^0_S$\ }
\def\KL{$K^0_L$\ }
\def\Kz{$K^0$\ }
\def\Kzb{$\overline{K^0}$\ }

\def\Repoe{$Re(\epsilon^{\prime}/\epsilon)$\ }
\def\KLpnn{$K^0_L\to\pi^0\nu\overline{\nu}$\ }
\def\Kppnn{$K^+\to\pi^+\nu\overline{\nu}$\ }
\newcommand{\be}{\begin{equation}}
\newcommand{\ee}{\end{equation}}
\newcommand{\bea}{\begin{eqnarray}}
\newcommand{\eea}{\end{eqnarray}}

\begin{document}

\title{ \vspace{1cm} Experiments with K-Meson Decays}


\author{T.\ K. Komatsubara\\
\\
High Energy Accelerator Research Organization (KEK), Japan }

\date{submitted on December 1, 2011\\
revised on March 19, 2012}

\maketitle

\begin{abstract}
Recent results and future prospects of
the particle physics experiments 
with neutral and charged K-meson (kaon) decays
are reviewed. 
Topics include 
  CP violation, 
  rare decays, 
  leptons in kaon decays, 
  tests of CPT and quantum mechanics,
  radiative decays,   
  hadrons in kaon decays, 
  basic observables,
  $V_{us}$ and CKM unitarity, 
and
  exotic searches.
Experimental techniques developed for the kaon decay experiments are discussed.

\end{abstract}

\section{Introduction
\label{sec:Intro}}

K-mesons (kaons) were discovered in 1947 in
``two (cloud-chamber) photographs (of cosmic ray showers)
containing forked tracks of a very striking character"~\cite{Rochester-Butler47}.
They were the first heavy-flavor particles,
which can be produced by strong interactions
but have only weak decays.
Through the decays 
they had major contributions~\cite{KaonPhysics} 
toward establishing the Standard Model (SM) of particle physics.
In the modern experiments, 
with millions of kaon decays per second by high-intensity accelerators,
searches and measurements with the sensitivity of $10^{-8} \sim 10^{-12}$ are 
performed and
the flavor parameters in and beyond the SM are studied. 

In this article~\footnote{
  The materials in my previous reviews~\cite{TKK01,TKK05,TKK10} are 
  expanded and updated.
}, 
recent results and future prospects of
kaon decay experiments (table~\ref{tab:kaonexp})  are reviewed.
Emphasis is placed on the achievements in the first decade of this century.
Theoretical kaon physics is not fully covered 
because comprehensive reviews are available
(e.g. \cite{1107.6001,1102.5650,FLAGwg,FlavLHC}).

\begin{table}
\begin{center}
\begin{minipage}[t]{16.5 cm}
\caption{
Kaon decay experiments being reviewed in this article.}
\label{tab:kaonexp}
\end{minipage}
  \begin{tabular}{l|lr|l|l}
 \hline
 Lab & \multicolumn{2}{|l|}{Accelerator} & Experiment & Kaon decay\\
 \hline
 KEK                                & KEK-PS       &(12 GeV)	& E246 $^{\surd}$          & \Kp at rest\\
                                         &                       &                        & E391a $^{\surd}$       & \KL\\
 BNL                                & AGS            &(25 GeV)	    & E949 $^{\surd}$\ ,\ E787 $^{\surd}$  & \Kp at rest\\
 KEK - JAEA$^{\rm a}$& J-PARC Main Ring   &(30 GeV)       &\KOTO   & \KL \\
                                         &                                      &                       &TREK               & \Kp at rest\\                
 IHEP, Protvino              & U-70           & (70 GeV)    & ISTRA+   $^{\surd}$                  & \Km in flight\\
                                          &                     &                     & OKA                                             & \Kpm in flight \\
 CERN                            & SPS	         &(400 GeV)   &NA48    $^{\surd}$                               & \KL, \KS\\
                                        &            	         &                     &NA48/1 $^{\surd}$                               & \KS\\
                                        &                       &                    &NA48/2 $^{\surd}$                                  & \Kpm in flight\\
                                        &                       &                    &NA62                                                      & \Kp in flight\\
 FNAL                             & Tevatron	&(800 GeV)  &KTeV $^{\surd}$    		& \KL, \KS\\
 INFN, Frascati             & \DAFNE       &($\sqrt{s}\sim{\rm 1.02GeV}$)    & KLOE$^{\surd}$                    & \KL+ \KS, \Kp + \Km \\
                                       &                          &                  & KLOE-2                                   & \KL + \KS, \Kp+ \Km \\

 \hline
  \end{tabular}
\begin{minipage}[t]{16.5 cm}
\vskip 0.5cm
\noindent
$^{\rm a}$ JAEA is the abbreviation of Japan Atomic Energy Agency. \\
$^{\surd}$ Data taking of the experiment is completed.
\end{minipage}
\end{center}
\end{table}

Reminder: the experimental upper limits in this article are 
at the 90\% confidence level (C.L.).

\clearpage

\section{CP Violation
\label{sec:CP}}

\subsection{\it Overview
\label{subsec:CP-Overview}}

The CP-violating $K^0_L\to\pi^+\pi^-$ decay was discovered~\cite{FC}, unexpectedly, in 1964
at the sensitivity of $10^{-3}$.
After the CP asymmetry in the \Kz - \Kzb 
mixing, with the parameter $\epsilon$, 
was established~\cite{F_RMP,C_RMP}, 
a long-standing problem has been its origin; 
the first question was 
whether it was due to the $\Delta S = 2$ {\em superweak} transition~\cite{Wolfenstein} or not. 
In 1973, Kobayashi and Maskawa~\cite{KM} accommodated 
CP violation in the electroweak theory
with six quarks
(and a single complex-phase in the mass-eigenstate mixing matrix
under the charged-current interactions).
The Kobayashi-Maskawa theory~\cite{K_RMP,M_RMP}, 
including  the prediction of 
{\em direct} CP violation in the decay process 
from the CP-odd component ($K_2$) to the CP-even state ($\pi\pi$), 
was verified by the observations of 
Time-reversal non-invariance (CPLEAR~\cite{CPLEAR} at CERN) and 
finally due to the determination of
$\epsilon^{\prime}/\epsilon$ (NA48 at CERN and KTeV at FNAL)
as well as the discoveries of top quark and  CP-violating B-meson decays. 

In the modern classification
(e.g. \cite{BTevRun2}) 
CP violation is grouped into three: 
in {\em mixing}, {\em decay}, and {\em interference between decays with and without mixing}
(table~\ref{tab:modernCP}). 
All of these have been extensively studied in the B Factory experiments.
Experimental studies in neutral-kaon decays have a long history, 
while the study of CP violation in the charged-kaon decay modes started recently. 
The CP-violating processes in  {\em mixing} and {\em decay} 
suffer from hadronic uncertainties.
A rare decay \KLpnn~\cite{litt89}, 
which will be discussed in Section~\ref{subsec:Rare-Kpnn},
is known to be a golden mode
in this category~\cite{GrossmanNir}
because the branching ratio can be calculated with very small 
theoretical-uncertainties in the SM as well as in its extensions.

The rest of this section is devoted to the kaon results 
of the CP violation in {\em decay} .

\begin{table}
\begin{center}
\begin{minipage}[t]{16.5 cm}
\caption{
Modern classification of CP violation.}
\label{tab:modernCP}
\end{minipage}
\begin{tabular}{l|c|l|c}
 \hline
   & example   & 
   hadronic uncertainties&
   kaon decays\\
 \hline
 mixing  & semileptonic decay 
                                                            & form factors  &                      \\
               &
 $\frac{\Gamma(\overline{M^0}(t)\ \to\ \ell^+ \nu X)\ -\ \Gamma({M}^0(t)\ \to\ \ell^- \overline{\nu} X)}
           {\Gamma(\overline{M^0}(t)\ \to\ \ell^+ \nu X)\ +\ \Gamma({M}^0(t)\ \to\ \ell^- \overline{\nu} X)}$
                                                             &                                           & \\             
               &                                           &                                           & $K^0\to \pi \mu \nu$,  $\pi e \nu$ $^{\surd}$\\
\hline
 decay   & hadronic decay
                                                             & hadronic& \\
   & 
 $M \to f$ \ vs.\ $\overline{M} \to \overline{f}$
                                                             & \ \ matrix elements, & \\

               &
 $\frac{\Gamma(M^-\ \to\ f)\ -\ \Gamma({M}^+\ \to\ \overline{f})}
           {\Gamma(M^-\ \to\ f)\ +\ \Gamma({M}^+\ \to\ \overline{f})}$
                                                           & strong phases & \\               
               &                                           &                                           &$K^0\to \pi \pi$ $^{\surd}$\\
               &                                           &                                           &\Kpm decays \\
 \hline
 interference   &
 $M^0 \to f_{CP}$ \ and\  $M^0 \to \overline{M}^0 \to  f_{CP}$
                                                             & form factors,  &   \\
  between decays &  
 $\frac{\Gamma(\overline{M}^0(t)\ \to\ f)\ -\ \Gamma({M}^0(t)\ \to\ f)}
           {\Gamma(\overline{M}^0(t)\ \to\ f)\ +\ \Gamma({M}^0(t)\ \to\ f)}$
                                                             & possible & \\
w/ and w/o mixing & & \ \ long-distance effects & \\
               &                                           &                                           &\KLpnn \\

 \hline
  \end{tabular}
\begin{minipage}[t]{16.5 cm}
\vskip 0.5cm
\noindent
$^{\surd}$ The CP violation has been observed.\\
$f_{CP}$: CP eigenstate
\end{minipage}
\end{center}
\end{table}

\subsection{\it CP Violation in decay
\label{subsec:CP-decay}}

\begin{figure}[tb]
\begin{center}
\begin{minipage}[t]{16.5cm}
\begin{center}
\epsfig{file=na48_detector_niels.eps,scale=0.50}
\vspace*{0.15cm}
\end{center}
\end{minipage}
\begin{minipage}[t]{16.5cm}
\begin{center}
\epsfig{file=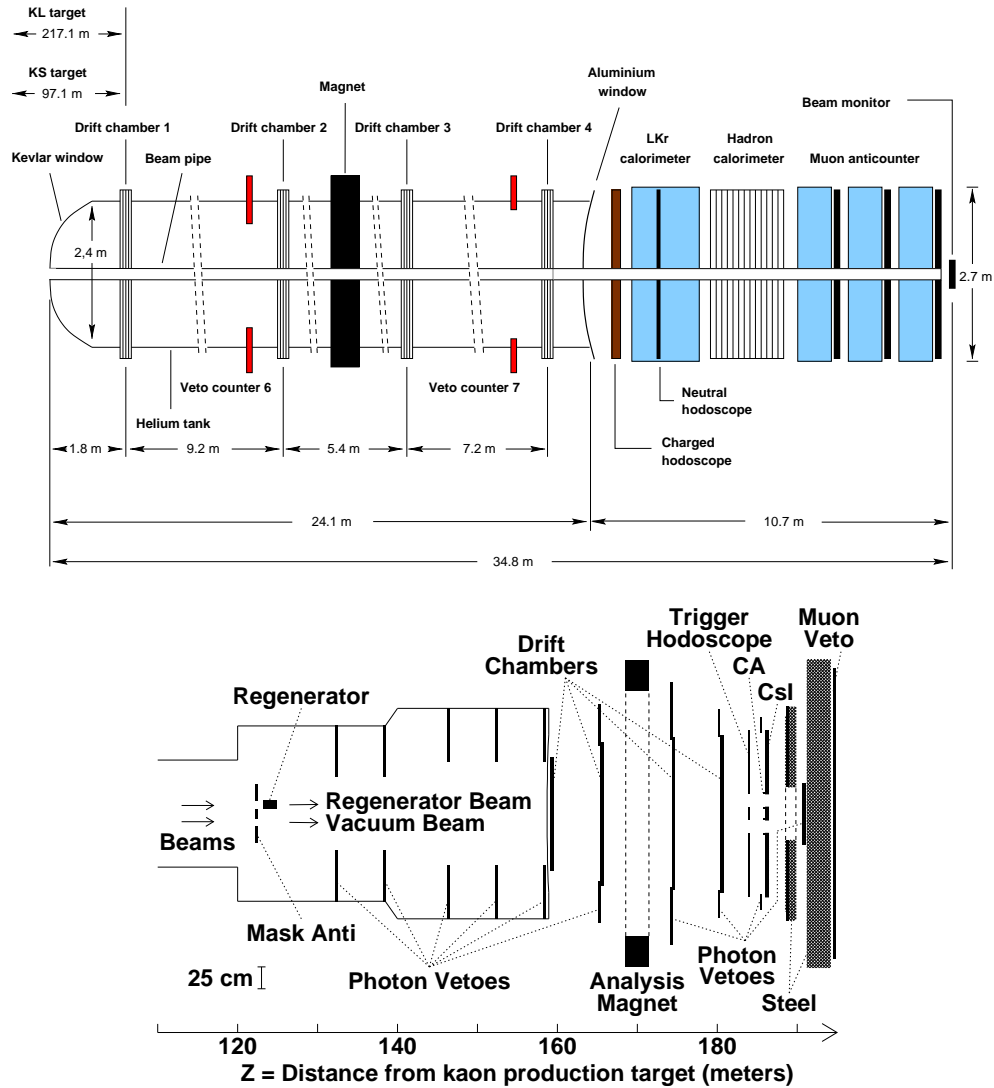,scale=0.50}
\vspace*{0.15cm}
\end{center}
\end{minipage}
\begin{minipage}[t]{16.5cm}
\caption{
The NA48 detector
with a magnetic spectrometer consisting of four drift chambers and a dipole magnet 
and a scintillator hodoscope for charged particles,
a liquid-Krypton calorimeter for photons,
and an iron-scintillator calorimeter and  a series of muon counters 
for charged-particle identification (top); 
the KTeV detector
with a regenerator, 
a magnetic spectrometer and a hodoscope, 
a cesium iodide calorimeter, 
and a muon counter (bottom).
Both detectors had a system of counters surrounding the fiducial region
to detect photons escaping the acceptance of the calorimeter.
Source: Figures taken from Refs.\cite{NA48-det2,KTeV-final}.
\label{Fig:epoedet}
}
\end{minipage}
\end{center}
\end{figure} 

\begin{figure}[tb]
\begin{center}
\begin{minipage}[t]{16.5cm}
\begin{center}
\epsfig{file=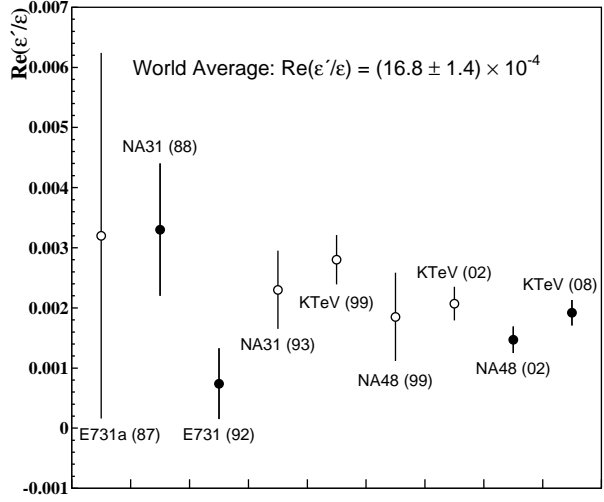,scale=0.40}
\vspace*{0.15cm}
\end{center}
\end{minipage}
\begin{minipage}[t]{16.5cm}
\caption{
Recent measurements of \Repoe .
Source: Figure taken from Ref.\cite{BWY_PTP}.
\label{Fig:epoeres}
}
\end{minipage}
\end{center}
\end{figure} 

 The NA48 collaboration at CERN and the KTeV collaboration at FNAL
published the final measurement of \Repoe
 as $ ( 14.7 \pm 2.2 ) \times 10^{-4}$~\cite{NA48-final} and 
$ ( 19.2 \pm 2.1 ) \times 10^{-4}$~\cite{KTeV-final}, respectively, 
where \Repoe is obtained from the double ratio of decay rates: 
\be
\frac{\Gamma(K^0_L\to\pi^0\pi^0)\ /\ \Gamma(K^0_S\to\pi^0\pi^0)}{\Gamma(K^0_L\to\pi^+\pi^-)\ /\ \Gamma(K^0_S\to\pi^+\pi^-)}
\ \approx\ 1 - 6\ Re(\epsilon^{\prime}/\epsilon)\;  \label{eq:dratio1}
\ee
or
\be
\frac{\Gamma(K^0_L\to\pi^+\pi^-)\ /\ \Gamma(K^0_S\to\pi^+\pi^-)}{\Gamma(K^0_L\to\pi^0\pi^0)\ /\ \Gamma(K^0_S\to\pi^0\pi^0)}
\ \approx\ 1 + 6\ Re(\epsilon^{\prime}/\epsilon)\; . \label{eq:dratio2}
\ee
The detectors for \Repoe~\cite{NA48-det,NA48-det2,KTeV-det}  are shown in Fig.~\ref{Fig:epoedet}.
All the four decay modes have to be measured simultaneously and with high precisions
(in particular to $K^0_L\to\pi^0\pi^0$); thus, 
a novel technique of double beams 
and a state-of-the-art electromagnetic calorimeter were developed.  
The NA48 experiment, 
shown in Fig.~\ref{Fig:epoedet} top,
used 
two target stations for kaon production~\footnote{
  A proton beam extracted from SPS  
  hit one target for \KL, and
  a lower-intensity proton beam,  
  which was selected by channelling the protons 
  not interacting in the \KL target in a bent silicon crystal, 
  was transported to a second target for \KS close to the detector.
  A time coincidence between an observed decay and a tagging counter,
  which was crossed by the low intensity proton beam on its way to the \KS target,
  was used to to identify the decay as coming from a \KS  rather than a \KL.}
and the liquid-Krypton (LKr) electromagnetic calorimeter
of 27 radiation-length ($X_0$) deep.
The KTeV experiment,
shown in Fig.~\ref{Fig:epoedet} bottom, 
used
side-by-side identical \KL beams with the regenerator alternating between them~\footnote{
  The regenerator created a coherent $|K^0_L> +\rho |K^0_S>$ state, 
  where $\rho$ is the regeneration amplitude.
   Most of the $K\to \pi\pi$ decay rate downstream of the regenerator 
   is from the \KS component. 
}
and the undoped cesium iodide (CsI) calorimeter of $27 X_0$ long. 
 Combining all the recent measurements of  \Repoe in Fig.~\ref{Fig:epoeres},
 the world average is
      $ ( 16.8 \pm 1.4 ) \times 10^{-4}$~\cite{BWY_PTP}; 
 it clearly demonstrates the existence of the CP violation in {\em decay} . 
 However, due to theoretical uncertainties in the hadronic matrix elements,  
 to get information on the SM and New Physics beyond it
 from \Repoe is difficult
 and remains to be a challenge to
 theoretical calculations.

\begin{figure}[tb]
\begin{center}
\begin{minipage}[t]{16.5cm}
\begin{center}
\epsfig{file=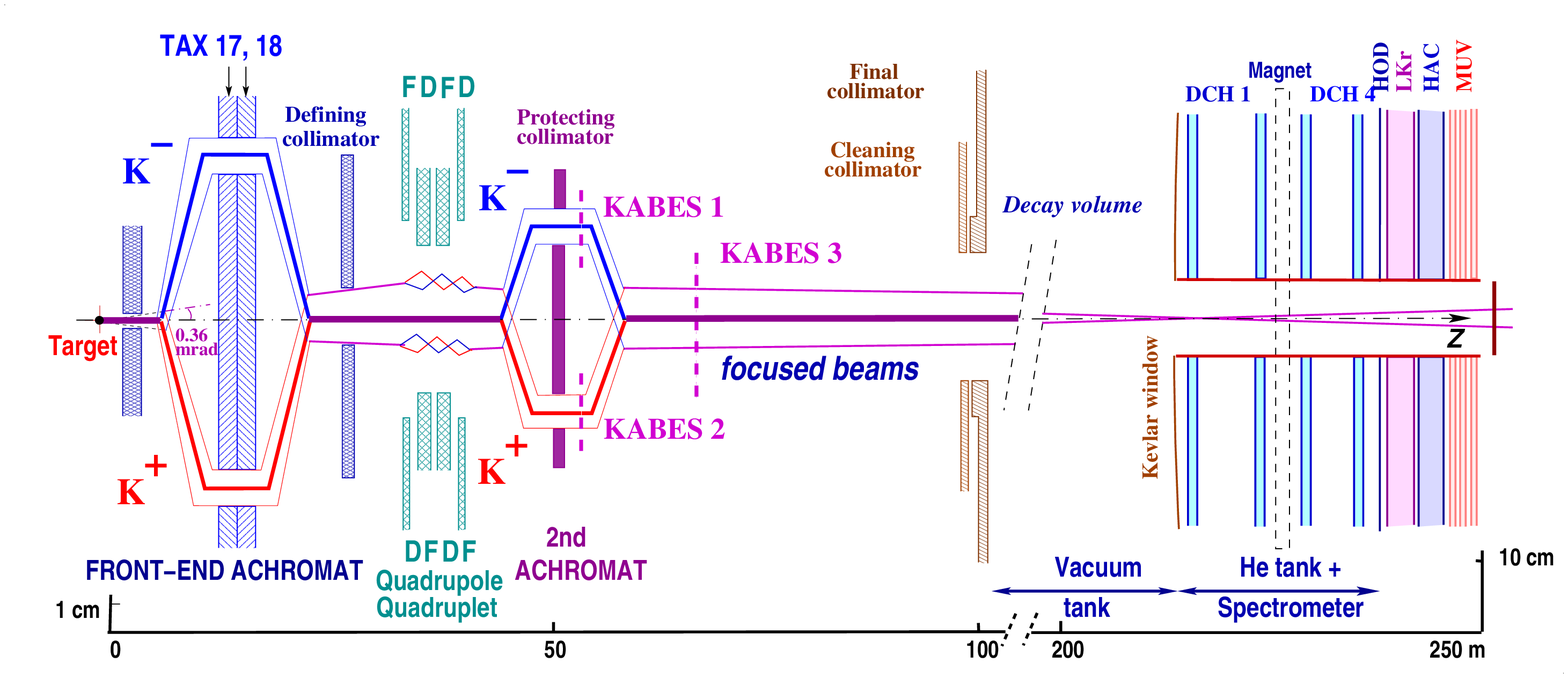,scale=0.45}
\end{center}
\end{minipage}
\begin{minipage}[t]{16.5cm}
\caption{
Schematic side view of the NA48/2 beam line, decay volume 
and detector.
TAX17, 18: motorized beam dump/collimators used to select
the momentum of the $K^+$ and $K^-$ beams; 
FDFD/DFDF: focusing magnets;
KABES 1-3: kaon beam spectrometer stations.
DCH 1-4: drift chambers;
HOD: hodoscope;
LKr: calorimeter; 
HAC: hadron calorimeter; 
MUV: muon veto.
Source: Figure taken from Ref.\cite{NA48Kp3}.
\label{Fig:na482beam}
}
\end{minipage}
\end{center}
\end{figure} 

 The NA48/2 collaboration at CERN performed charge asymmetry measurements
 with the simultaneous \Kp and \Km beams
 (Fig.~\ref{Fig:na482beam}), 
 overlapping in space,  of $60\pm 3$ GeV/$c$ in 2003 and 2004.
 The $K^{\pm}\to \pi^{\pm}\pi\pi$ matrix element squared (Dalitz-plot density) is 
 parametrized~\cite{Dalitzdensity} by a polynomial expansion
 with two Lorentz-invariant kinematic variables $u$ and $v$: 
 \be
 | M(u,v) |^2\ \propto\  1\ +\ g\cdot u\ +\ h \cdot u^2\ +\ k \cdot v^2\ , \label{eq:kp3sq}
\ee 
 where the coefficients $g$, $h$, $k$ are called as the linear and quadratic {\em slope parameters}.
 The linear slope parameters, $g^+$ and $g^-$,  describe the decays of $K^+$ and $K^-$, respectively,
 and the slope asymmetry $A_g\ =\ (g^+ - g^-)/(g^+ + g^-)$ is a manifestation of 
 CP violation in {\em decay}. 
  The asymmetries of  the $K^{\pm} \to \pi^{\pm} \pi^+ \pi^-$ decay $A_g^c$
  and the $K^{\pm} \to \pi^{\pm} \pi^0 \pi^0$ decay $A_g^n$
  were measured to be 
      $A_g^c = ( -1.5 \pm 2.2 ) \times 10^{-4}$ 
      and
      $A_g^n = ( 1.8 \pm 1.8) \times 10^{-4}$
 with the data sets of 4G and 0.1G events, respectively~\cite{NA48Kp3}.
 The SM expectation is in $10^{-5}\sim 10^{-6}$, and 
 no evidence for CP violation in {\em decay} 
 was observed in the $K^{\pm}$ decays at the level of $2\times 10^{-4}$. 
 The NA48/2 experiment also measured the asymmetries of $K^+$ and $K^-$ decay-rates  
 in $K^{\pm} \to \pi^{\pm} e^{+} e^{-}$,
 $K^{\pm} \to \pi^{\pm} \mu^{+} \mu^{-}$
 and
  $K^{\pm} \to \pi^{\pm} \pi^0 \gamma$
 to be 
      $ ( -2.2 \pm 1.5(stat.) \pm 0.6 (syst.) ) \times 10^{-2}$~\cite{NA48-piee}, 
      $ (   1.2 \pm 2.3 ) \times 10^{-2}$~\cite{NA48-pimm}
      and
      $ ( 0.0 \pm 1.0(stat.) \pm 0.6 (syst.) ) \times 10^{-3}$~\cite{NA48-ppg}
 and set the upper limits of $2.1$\%, $2.9$\% and $0.15\%$, respectively. 

No new plan for the \Repoe and $A_g$ measurements  is being  proposed.

\clearpage

\section{Rare decays
\label{sec:Rare}}

\subsection{\it Overview
\label{subsec:Rare-Overview}}

A famous contribution of kaon physics to the SM was
the absence of the decays due to 
Flavor-Changing Neutral Current (FCNC), 
$K^0_L\to\mu^+\mu^-$, $K^+\to\pi^+ e^+ e^-$ and  $K^+\to\pi^+\nu\bar{\nu}$,
in 1960's. 
The Glashow-Iliopoulos-Maiani (GIM) mechanism~\cite{GIM70}
explained it~\cite{GL74} with the the unitarity matrix 
for mass-eigenstate mixing
in advance of the charm-quark discovery in 1974. 
The next theoretical milestone was 
the Inami-Lim loop functions~\cite{IL81} in 1981
for FCNC processes with heavy quark and leptons. 
In addition to the continual searches for decays
at higher levels than the SM predictions, 
the anticipation of detecting rare kaon decays  
in the SM range got more and more realistic 
with the rise of the top-quark mass. 
The evidence for the $K^+\to\pi^+\nu\bar{\nu}$ decay~\cite{E787evidence} 
by the E787 collaboration at BNL in 1997
opened a new era of testing the SM by measuring rare processes.

\begin{figure}[tb]
\begin{center}
\begin{minipage}[t]{16.5cm}
\begin{center}
\epsfig{file=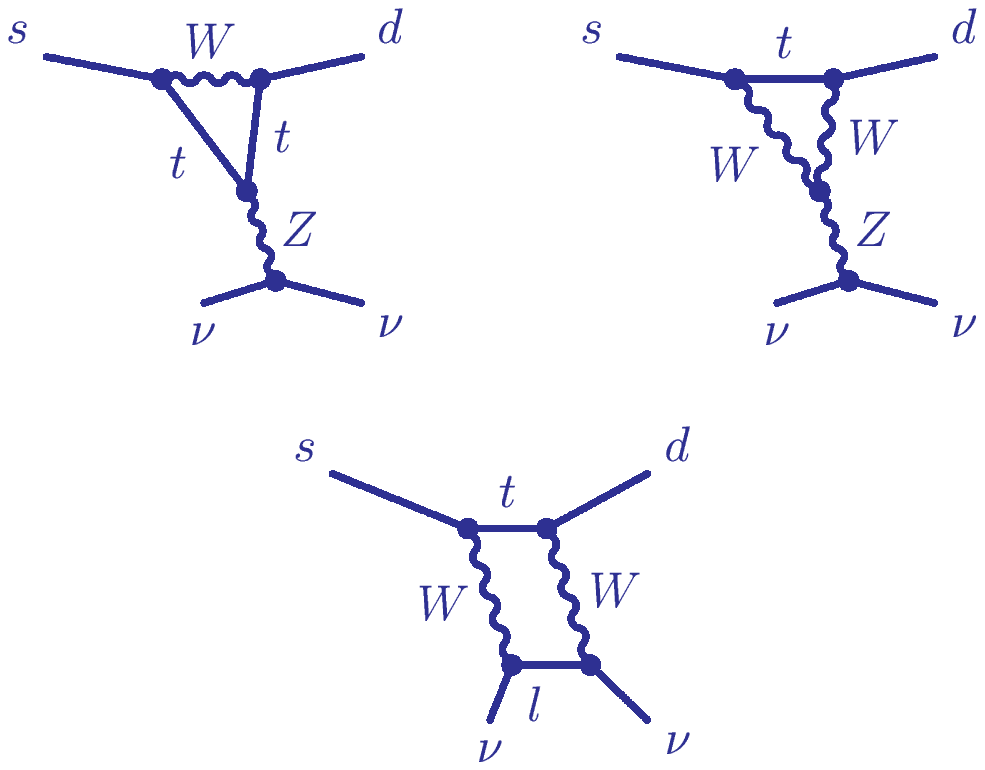,scale=0.65}
\hspace*{1.5cm}
\epsfig{file=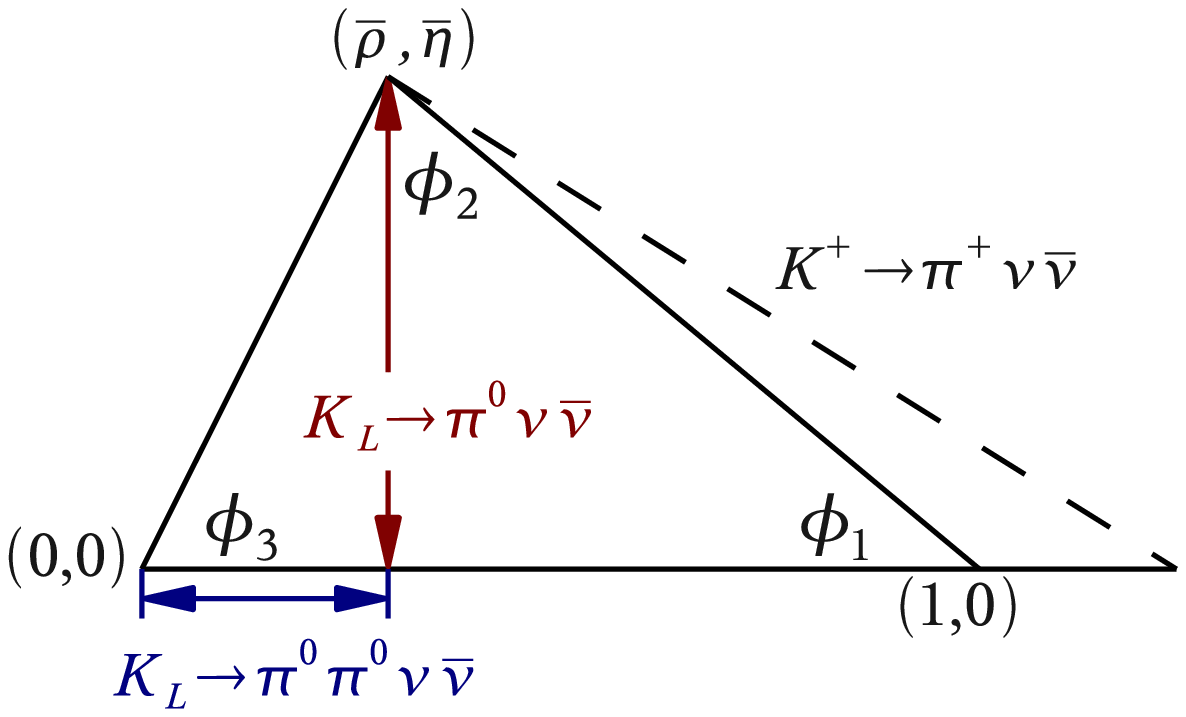,scale=0.65}
\end{center}
\end{minipage}
\begin{minipage}[t]{16.5cm}
\caption{SM diagrams for $K\to\pi\nu\bar{\nu}$ (left);
kaon unitarity triangle in the ($\rho$,$\eta$) plane (right).
\label{Fig:pnndiag}
}
\end{minipage}
\end{center}
\end{figure} 

The FCNC process in kaon decays
is strange-quark to down-quark transition 
and is induced in the SM by the electroweak loop effects
as Penguin and Box diagrams (Fig.~\ref{Fig:pnndiag} left).
The top-quark in the loops dominates the transition 
because of its heavy mass, and the quantity: 
\be
   \lambda_t \ \equiv\  
   V_{ts}^{*} \cdot V_{td} \ =\   
      - A^{2} \lambda^5 \cdot (1 - \rho - i \eta) \ =\ 
      - |V_{cb}|^{2} \cdot \lambda \cdot (1 - \rho - i \eta)\ ,
\label{eq:vtsvtd}
\ee
where 
$\lambda \equiv |V_{us}| \simeq 0.225$ (to be discussed in Section~\ref{sec:Vus}), 
is measured.
$\lambda$, $A$, $\rho$, and $\eta$ 
are
the Wolfenstein parametrization~\cite{WolfensteinP}
of the Cabibbo-Kobayashi-Maskawa(CKM) matrix~\cite{Cabibbo,KM}.
The decays are rare
due to $|V_{cb}|^{2} \cdot \lambda$, and
are precious because the important parameters $\rho$ and $\eta$
can be determined from them. 
The decay amplitude of \KL is 
a superposition of the amplitudes of \Kz\ and \Kzb 
and is proportional to $\eta$ ; 
thus, observation of a rare \KL decay to a CP-even state
is a new evidence for CP violation. 

We start with \KLpnn and
 the charged counterpart \Kppnn, 
 because measurement of their branching ratios
is currently the main issue of kaon physics~\cite{SozziCKM2010}.

\subsection{\it $K\to \pi\nu\bar{\nu}$
\label{subsec:Rare-Kpnn}}

The branching ratios 
of the $K\to\pi\nu\bar{\nu}$ decays~\cite{BUS08} 
are predicted in the SM, to an exceptionally high degree of precision,
as~\cite{BGS11} 
\bea
     B(K^0_L \to \pi^0\nu\bar{\nu}) & = & 
      \kappa_{L} \left( \rm{Im} \frac{\lambda_{t}}{\lambda^5} \cdot X_{t}  \right)^2 
\label{eq:pi0nn}
\eea
$= 2.43\ (39)\ (06) \times 10^{-11}\ $ and
\bea
     B(K^+ \to \pi^+\nu\bar{\nu}(\gamma)) & = &
     \kappa_{+}\ (1+\Delta_{\rm{EM}})\ 
     \left[
     \left( \rm{Im} \frac{\lambda_{t}}{\lambda^5} \cdot X_{t}  \right)^2
     + 
     \left( 
           \rm{Re} \frac{\lambda_{t}}{\lambda^5}\cdot X_{t}  
       \ +\  
          \rm{Re} \frac{\lambda_{c}}{\lambda}\cdot (P_c + \delta P_{c,u})
    \right)^2
     \right] 
\label{eq:pipnn}
\eea
$= 7.81\ (75)\ (29) \times 10^{-11}\ $.
The first error summarizes the parametric uncertainties
dominated by $|V_{cb}|$, $\rho$ and $\eta$, and
the second error summarizes  the remaining theoretical uncertainties.
Long-distance contributions to $K\to\pi\nu\bar{\nu}$  are small; 
the hadronic matrix elements contained in the parameters $\kappa_{L}$ and $\kappa_{+}$
can be extracted
from the $K_{\ell 3}$ decays (Section~\ref{sec:Basic}), and
the long-distance QED corrections are denoted by $\Delta_{\rm EM}$~\cite{Mescia-Smith07}.
${\lambda_{t}}/{\lambda^5}$ and ${\lambda_{c}}/{\lambda}$, 
where  $\lambda_c \equiv V_{cs}^{*} \cdot V_{cd}$, 
represent the CKM-matrix parameters in $K\to\pi\nu\bar{\nu}$
and are in $O(\lambda^0)$.
The decay amplitudes of \KLpnn and \Kppnn
are proportional to the height (and thus the area) and 
the length of a side
of the CKM unitarity triangle, respectively,
giving access to the parameters $\rho$ and $\eta$
as shown in Fig.~\ref{Fig:pnndiag} right.
The function $X_{t}$ is the top-quark contribution to  $K\to\pi\nu\bar{\nu}$; 
the complete two-loop electroweak corrections to $X_{t}$~\cite{BGS11} 
as well as 
the next-to-next-to-leading order QCD corrections~\cite{BGHN06}
and the QED and electroweak corrections~\cite{Brod-Gorbahn08}
to the charm quark contribution to
\Kppnn, parametrized by 
$P_c$ and $\delta P_{c,u}$, 
have been calculated.

\begin{figure}[tb]
\begin{center}
\begin{minipage}[t]{16.5cm}
\begin{center}
\epsfig{file=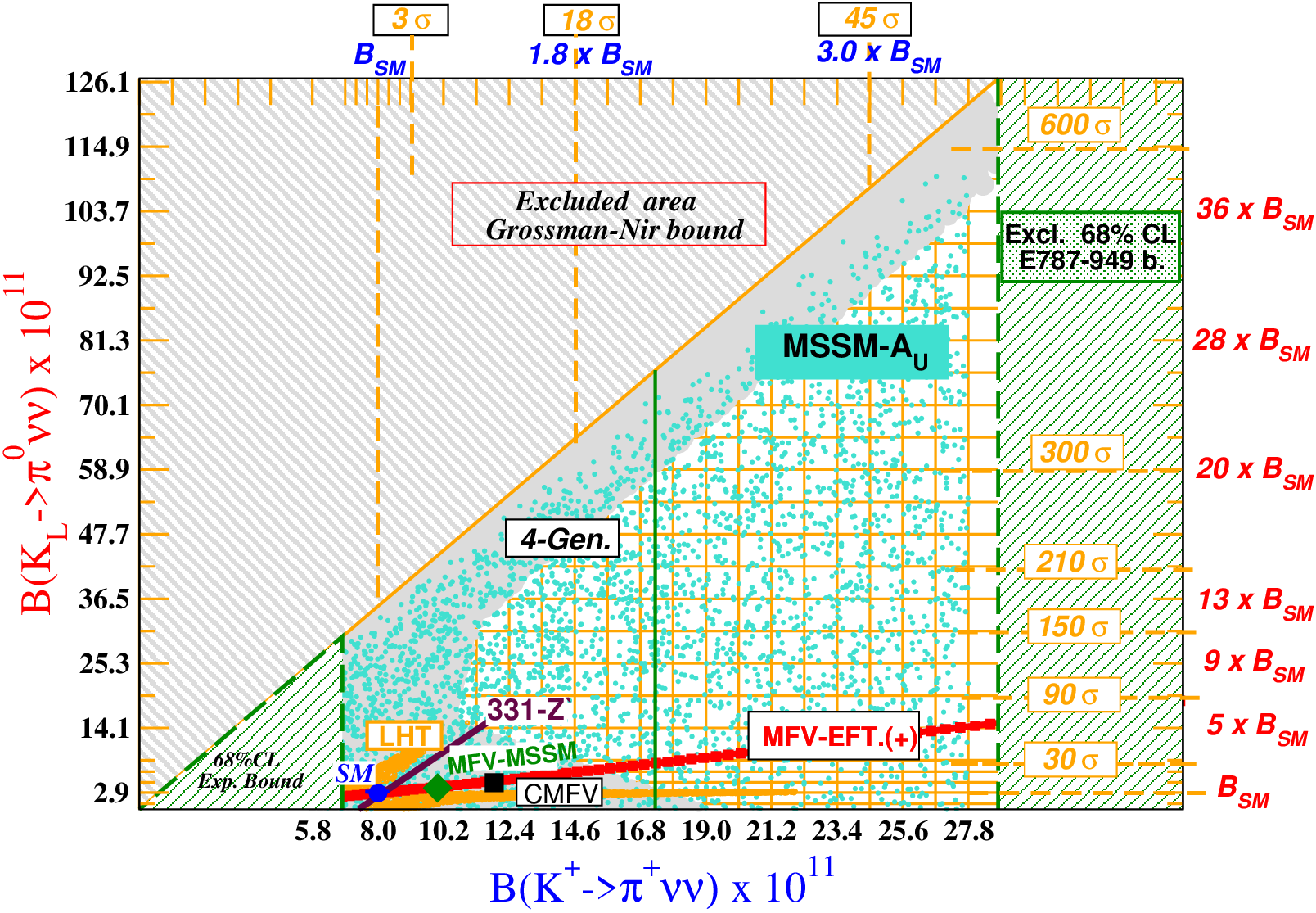,scale=0.50}
\end{center}
\end{minipage}
\begin{minipage}[t]{16.5cm}
\caption{
\KLpnn branching ratio versus \Kppnn branching ratio plot
of the theoretical predictions
in the Standard Model ($SM$) and in its extensions such as
the minimal flavor violation (MFV),
constrained minimal flavor violation (CMFV), 
minimal Supersymmetric Standard Model (MSSM), 
littlest Higgs model with T-parity (LHT),
minimal 3-3-1 model (331-Z$^{\prime}$),
and four-generation model ($4$-$Gen.$).
Excluded areas 
by the measured \Kppnn branching ratio
and the model-independent bound due to their isospin relation are also shown in the plot.
Source: Figure taken from Ref.\cite{Krarehtml}.
\label{Fig:Krare_BSM_new}
}
\end{minipage}
\end{center}
\end{figure}

New Physics could affect 
these branching ratios~\cite{BBIL06,Krarehtml,BurasMLV,1012.3893,1104.0826} 
and, by the measurement, 
the flavor structure in New Physics 
(operators and phases in the interactions of new particles) 
can be studied. 
The $K\to\pi\nu\bar{\nu}$ branching ratios beyond the Standard Model
are presented in Fig.~\ref{Fig:Krare_BSM_new}.
A model-independent bound 
$B(K^0_L \to \pi^0\nu\bar{\nu}) < 4.4\times B(K^+ \to \pi^+\nu\bar{\nu})$, 
called the Grossman-Nir bound~\cite{GrossmanNir}, 
can be extracted from their isospin relation.

The signature of $K\to\pi\nu\bar{\nu}$ 
is a kaon decay into a pion plus {\em nothing}.
Background rejection is essential in the experiments, 
and {\em blind analysis} techniques have been developed and refined 
to achieve a high level of confidence in the background measurements.
To verify {\em nothing}, 
hermetic extra-particle detection
by photon and charged-particle detectors, 
called the {\em veto} as a jargon in the experiments, 
is imposed to the hits in coincidence with the pion time 
and
with the visible-energy threshold less than a few MeV.
Tight veto requirements are indispensable
in order to achieve a low detection-inefficiency $< 10^{-3}\sim 10^{-4}$; 
good timing resolution for low energy hits is therefore  
essential to avoid acceptance loss due to accidental hits
in the environment of high-intensity beam.

\begin{figure}[tb]
\begin{center}
\begin{minipage}[t]{16.5cm}
\begin{center}
\epsfig{file=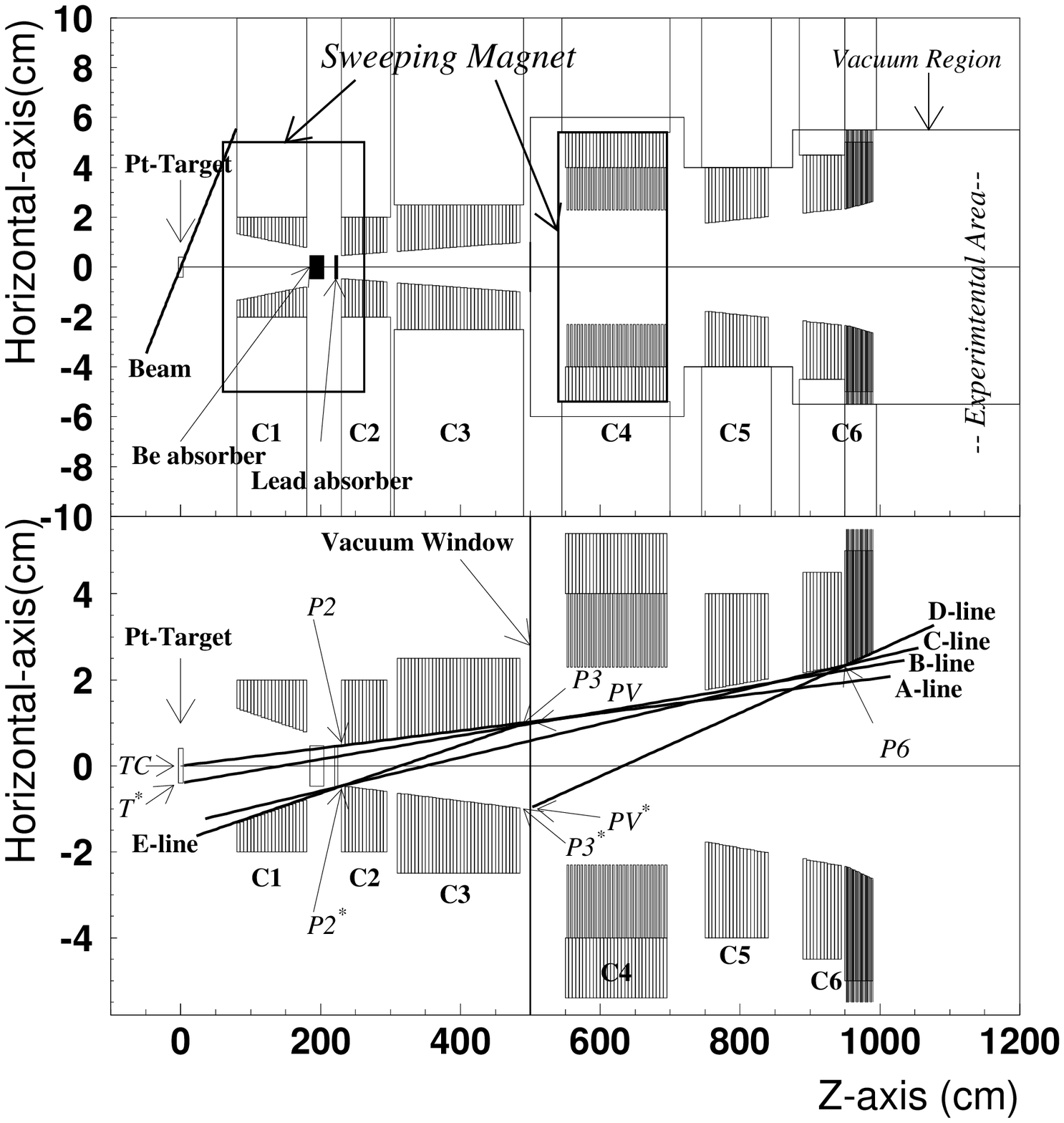,scale=0.45}
\vspace*{1.0cm}
\end{center}
\end{minipage}
\begin{minipage}[t]{16.5cm}
\begin{center}
\epsfig{file=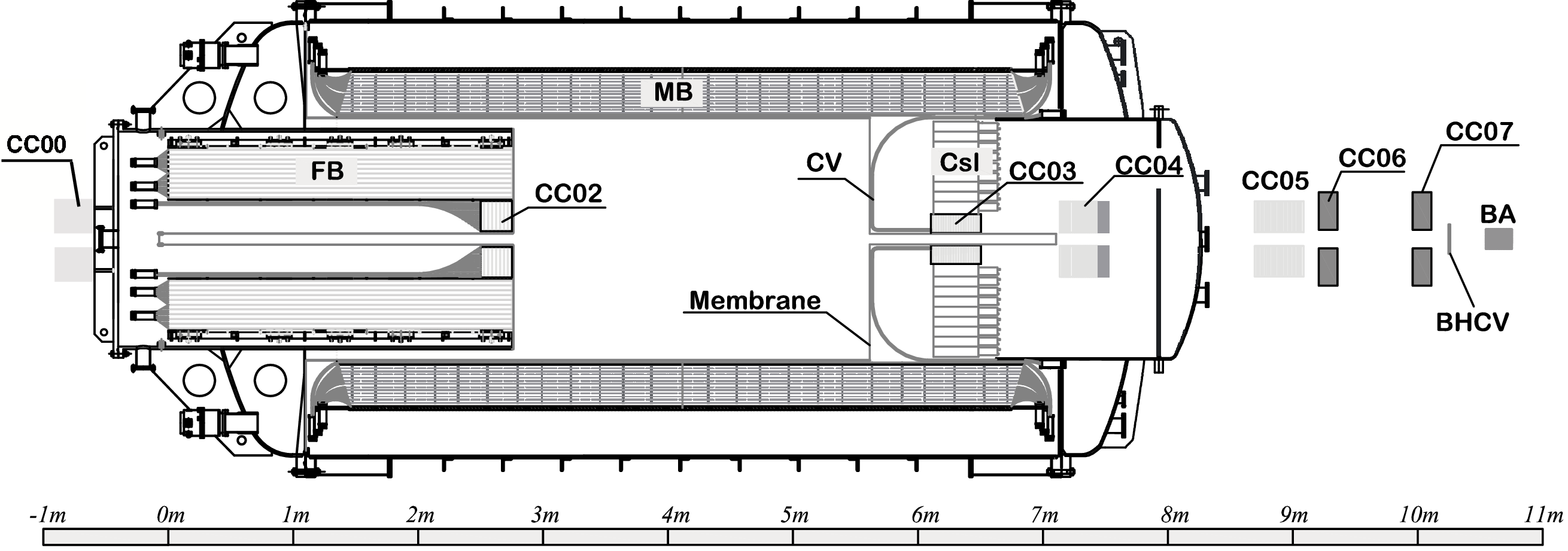,scale=0.50,angle=270}
\vspace*{0.15cm}
\end{center}
\end{minipage}
\begin{minipage}[t]{16.5cm}
\caption{
Schematic horizontal views of the neutral beam line for E391a:
the arrangements of the components
including six collimators C1-C6, 
and the collimation scheme (top); 
cross-sectional view of the E391a detector (bottom).
CsI: electromagnetic calorimeter;
CV: charged-particle counter;
MB and FB: main-barrel and front-barrel photon counters; 
CC00,CC02-CC07: collar-shaped photon counters;
BHCV and BA: beam hole charged-particle and photon counters.
Source: Figures taken from Refs.\cite{pencil,E391afinal}.
\label{Fig:E391adet}
}
\end{minipage}
\end{center}
\end{figure} 

The E391a collaboration at KEK performed
the first dedicated search for the \KLpnn decay.
A well-collimated, small-diameter neutral beam 
(called a {\em pencil} beam~\cite{pencil}, Fig.~\ref{Fig:E391adet} top) 
was designed and constructed. 
The $K^0_L$ beam, whose momentum peaked at 2 GeV/$c$, 
was produced by the 12-GeV protons of KEK-PS.
The E391a detector is shown in Fig.~\ref{Fig:E391adet} bottom.
The energy and position of the two photons from $\pi^0$ decays were measured 
by a downstream electromagnetic calorimeter
consisting of 576 undoped-CsI crystals 
in the size of 7.0$\times$7.0$\times$30 cm$^3$ (i.e. 16$X_0$ long)~\cite{E391aCal}.
Assuming that the invariant mass of two photons was equal to the $\pi^0$ mass
and that the decay vertex was on the beam axis, 
the $K^0_L$-decay vertex position, ${\rm Z_{vtx}}$, 
and the transverse momentum of $\pi^0$, ${\rm P_{T}}$, were determined. 
A $\pi^0$ with a large transverse momentum 
(${\rm P_{T}}$ $\ge$ 0.12 GeV/$c$) 
was the signal.
In order to suppress the major background from $K^0_L\to\pi^0\pi^0$,
the remaining part of the calorimeter not hit by the two photons 
as well as all the other detector subsystems 
covering the decay region~\cite{E391aBarrel,E391aBA}
were used as a veto.
The beam line and the collimation scheme were designed carefully 
to minimize the beam halo (mostly neutrons), 
which could interact with the counters near the beam 
and produce $\pi^0$'s and $\eta$'s.

\begin{figure}[tb]
\begin{center}
\begin{minipage}[t]{16.5cm}
\begin{center}
\epsfig{file=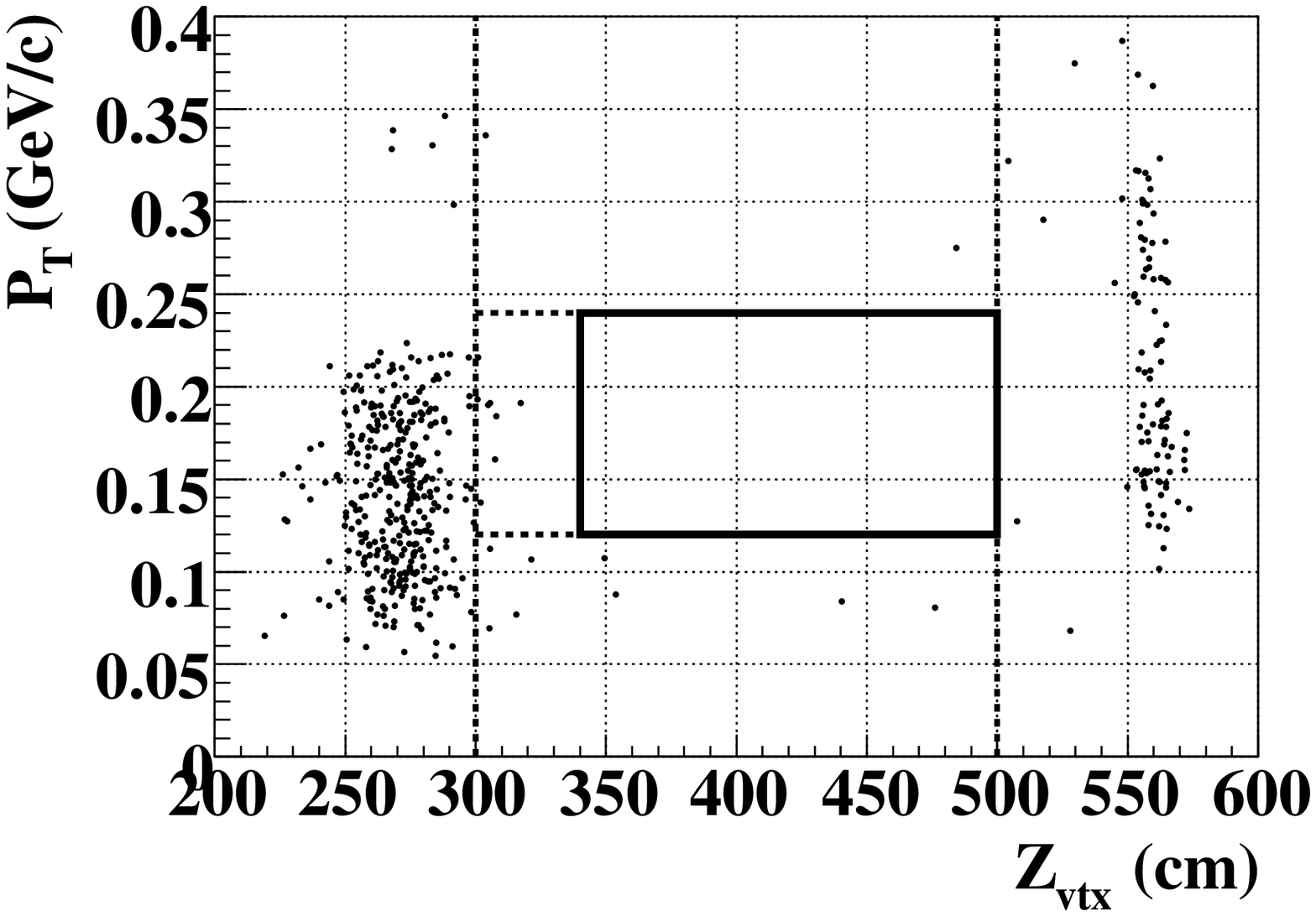,scale=0.55}
\end{center}
\end{minipage}
\begin{minipage}[t]{16.5cm}
\caption{
E391a result on $K^0_L \to \pi^0\nu\bar{\nu}$.
Source: Figure taken from Ref.\cite{E391afinal}.
\label{Fig:kpnnE391a}
}
\end{minipage}
\end{center}
\end{figure}

In the final results from E391a on \KLpnn~\cite{E391afinal}, 
the single event sensitivity was $1.11\times 10^{-8}$ and no events were observed
inside the signal region (Fig.~\ref{Fig:kpnnE391a}).
The upper limit on $B(K^0_L \to \pi^0\nu\bar{\nu}) $ was set to be
$2.6\times 10^{-8}$.
The E391a experiment has improved the limit from previous experiments
($5.9\times 10^{-7}$~\cite{KTeV-pnnDalitz} 
using $\pi^0\to e^+ e^-\gamma$)~\footnote{
   The improvement from the limit using $\pi^0\to\gamma\gamma$, 
   $1.6\times 10^{-6}$~\cite{KTeV-pnngg}, is a factor of 62.}
by a factor of 23.
E391a also obtained an upper limit of 
$8.1 \times 10^{-7}$~\cite{E391appnn}
for the branching ratio of the $K^0_L\to\pi^0\pi^0\nu\bar{\nu}$ decay, 
which is a CP conserving process and is sensitive to 
$\rm{Re} \lambda_{t}$ (Fig.~\ref{Fig:pnndiag} right).

\begin{figure}[tb]
\begin{center}
\begin{minipage}[t]{16.5cm}
\begin{center}
\epsfig{file=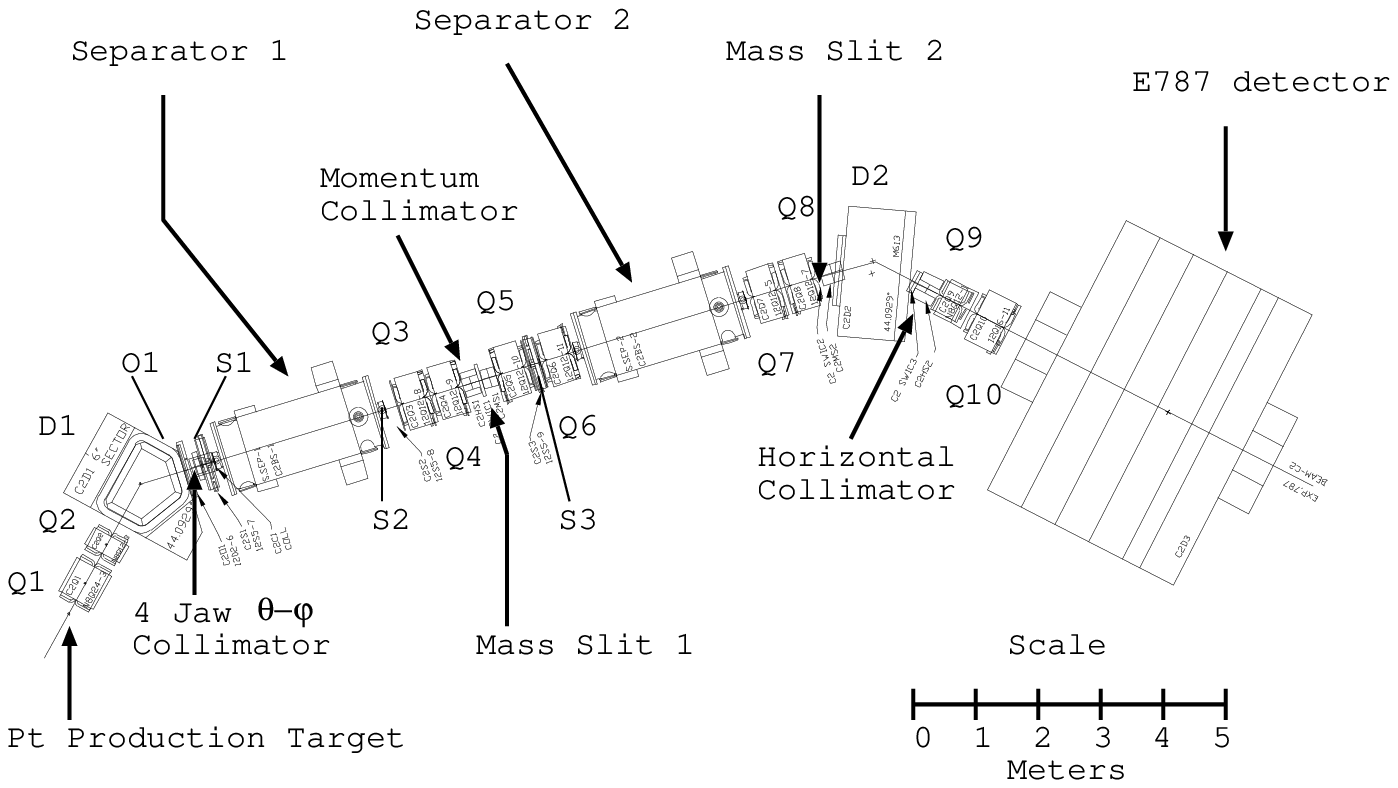,scale=0.90}
\end{center}
\end{minipage}
\begin{minipage}[t]{16.5cm}
\begin{center}
\epsfig{file=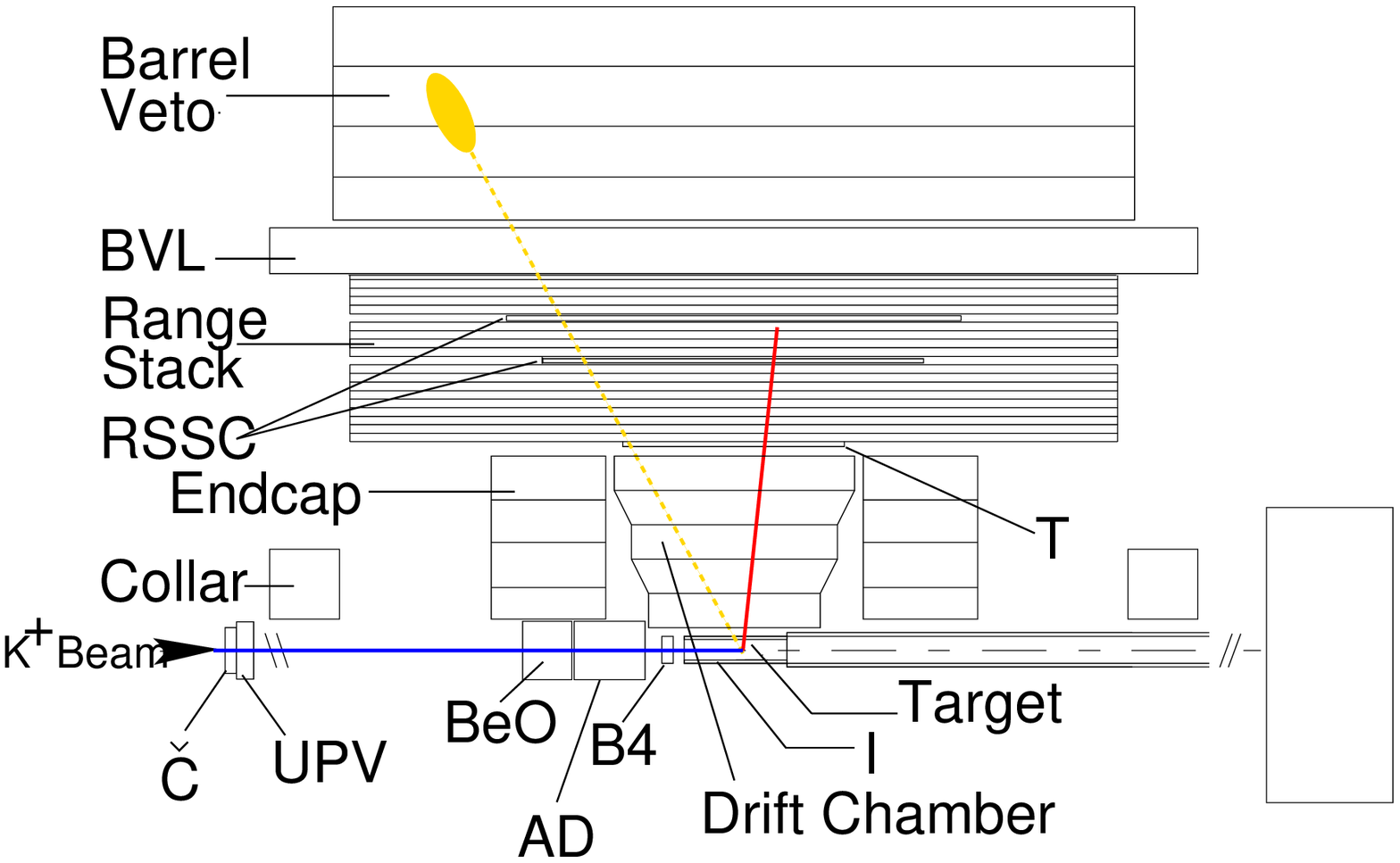,scale=0.40}
\hspace*{1.0cm}
\epsfig{file=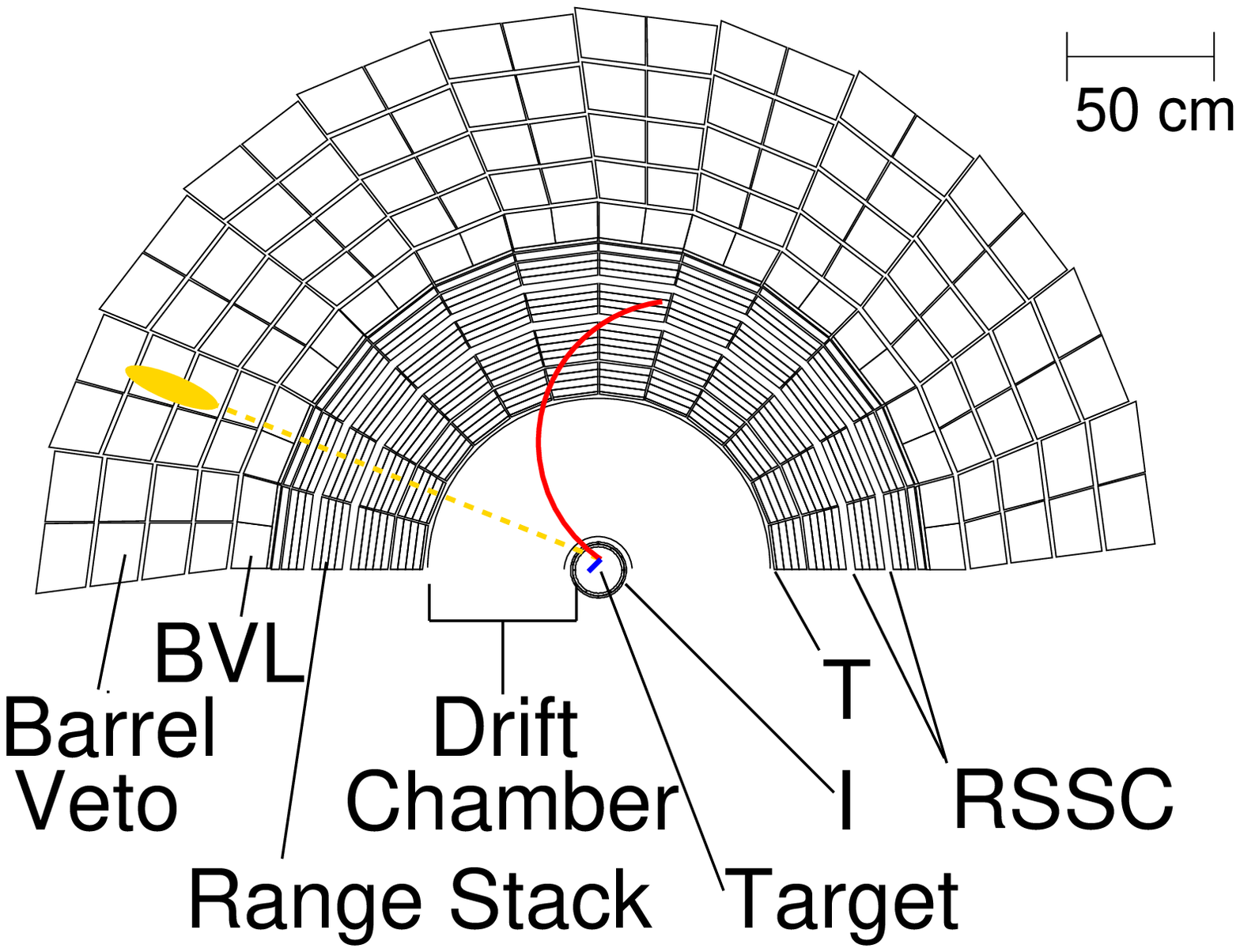,scale=0.40}
\vspace*{0.15cm}
\end{center}
\end{minipage}
\begin{minipage}[t]{16.5cm}
\caption{
Layout of LESB III of AGS for the E949/E787 experiment
with two dipole (D1,D2),
ten quadrupole (Q1-Q10),
three sextupole (S1-S3)
and one octupole (O1) magnets (top);
schematic side (bottom left) and end (bottom right) views 
of the upper half of the E949 detector.
The outgoing charged pion and one photon from $\pi^0\to\gamma\gamma$ decay
are illustrated.
\v{C}: \v{C}erenkov counter;
B4: energy-loss counters; 
I and T: inner and outer trigger scintillation counters;
RSSC: Range Stack straw-tube tracking chambers;
BVL: barrel-veto-liner photon counter;
UPV: upstream photon-veto counter; 
AD: active degrader.
Source: Figures taken from Refs.\cite{LESB3,E949final}.
\label{Fig:E949det}
}
\end{minipage}
\end{center}
\end{figure} 

The E949 and E787 collaborations at BNL measured 
the charged track emanating from \Kppnn
decaying at rest in the stopping target.
The Low Energy Separated Beam line (LESB III~\cite{LESB3}, Fig.~\ref{Fig:E949det} top) of AGS
transported \Kp's with momentum 0.7 GeV/$c$ to the detector (Fig.~\ref{Fig:E949det} bottom). 
Pion contamination to the incident $K^+$ beam
was reduced, to a $K^+ : \pi^+$ ratio of $4:1$,
by two stages of electrostatic particle separation in LESB III
to prevent that
scattered beam pions would contribute to the background.
Charged-particle detectors for 
measurement of the $\pi^+$ properties
were located in the central region of the detector
and were surrounded by hermetic photon detectors.
The $\pi^+$ momentum ($P_{\pi^+}$) from \Kppnn is less than 0.227 GeV/$c$, 
while the major background sources of 
$K^+\to\pi^+\pi^0$ and $K^+\to\mu^+\nu$
are two-body decays and 
have monochromatic momentum of 0.205 GeV/$c$ and 0.236 GeV/$c$, respectively.
Two signal regions
with the $\pi^+$ momentum
above and below the peak from  $K^+\to\pi^+\pi^0$ were adopted.
Redundant kinematic measurement and 
$\mu^+$ rejection were employed; 
the latter was crucial
in the trigger as well as in the offline analysis
because the $K^+\to\mu^+\nu$ background had the same topology as the signal.

\begin{figure}[tb]
\begin{center}
\begin{minipage}[t]{16.5cm}
\begin{center}
\epsfig{file=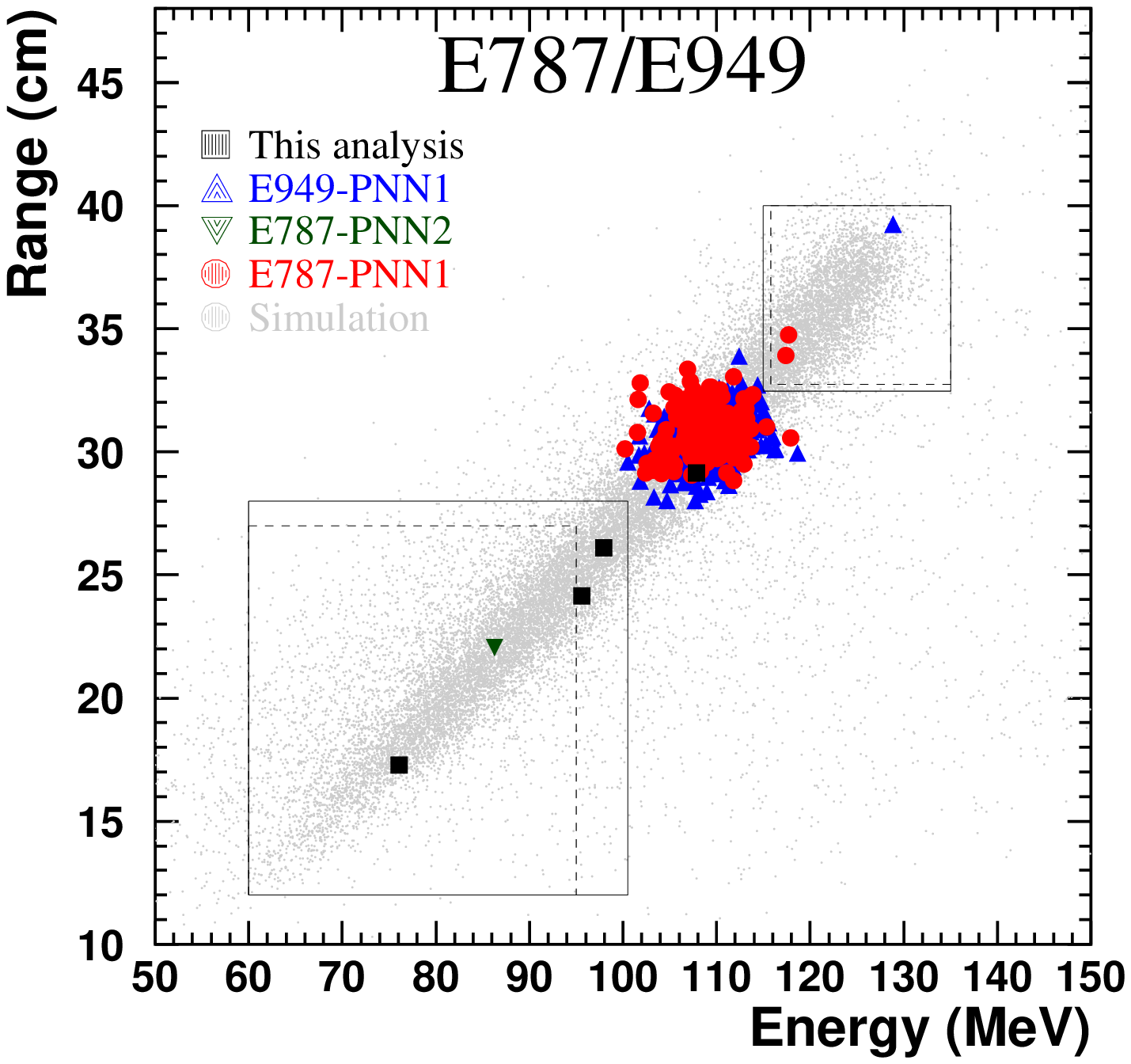,scale=0.55}
\end{center}
\end{minipage}
\begin{minipage}[t]{16.5cm}
\caption{
E949/E787 result on $K^+ \to \pi^+\nu\bar{\nu}$: 
$\pi^+$ kinetic energy versus range plot of all events passing all other selection criteria.
The solid (dashed) lines represent the limits of the PNN1/PNN2 signal regions
for the E949 (E787) analyses. 
The light gray points are simulated \Kppnn events that were accepted by the trigger.
Source: Figure taken from Ref.\cite{E949final}.
\label{Fig:kpnnE949}
}
\end{minipage}
\end{center}
\end{figure}

The E949 experiment observed 
one \Kppnn event in the kinematic region 
$0.211 < P_{\pi^+} < 0.229$ GeV/$c$ (PNN1)~\cite{pnn1final}
and three events in the region 
$0.140 < P_{\pi^+} < 0.199$ GeV/$c$ (PNN2)~\cite{E949final}.
Combining the results 
with the observation of two events in PNN1 and one event in PNN2
by the predecessor E787 experiment,
a branching ratio of 
$B(K^+ \to \pi^+\nu\bar{\nu}) = (1.73^{+1.15}_{-1.05})\times 10^{-10}$
(Fig.~\ref{Fig:kpnnE949})~\cite{E949final}
was obtained 
and was consistent with the Standard Model prediction.
The upper limit $B(K^+ \to \pi^+\nu\bar{\nu}) < 3.35 \times 10^{-10}$
was also determined
and can be used to calculate the Grossman-Nir bound
on $B(K^0_L \to \pi^0\nu\bar{\nu})$
as $1.46 \times 10^{-9}$,
which is smaller than the limit by E391a. 
The E787 experiment also obtained an upper limit of 
$4.3 \times 10^{-5}$~\cite{E787ppnn}
for the branching ratio of the $K^+\to\pi^+\pi^0\nu\bar{\nu}$ decay.

\begin{figure}[tb]
\begin{center}
\begin{minipage}[t]{16.5cm}
\begin{center}
\epsfig{file=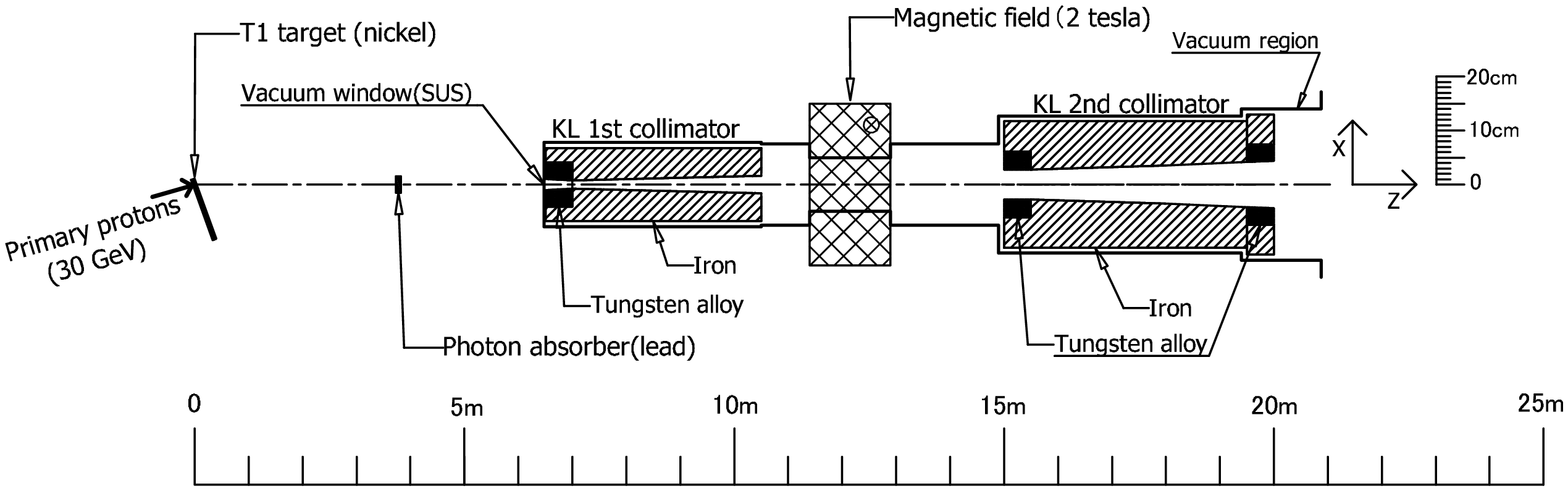,scale=0.60, angle=90}
\end{center}
\end{minipage}
\begin{minipage}[t]{16.5cm}
\caption{
Schematic plan view of the neutral beam line for \KOTO .
Source: Figure taken from Ref.\cite{KOTObeam}.
\label{Fig:KOTObeam}
}
\end{minipage}
\end{center}
\end{figure}

The next generation of \KLpnn
is the E14 \KOTO experiment~\cite{KOTOweb,KOTOnanjo,KOTOwatanabe} 
at the new high-intensity proton accelerator facility 
J-PARC (Japan Proton Accelerator Research Complex)~\cite{JPARC}. 
The accelerators consist of a Linac, 3-GeV Rapid Cycle Synchrotron 
and Main Ring, 
and the physics program
with slow and fast beam-extractions of 30 GeV protons has started~\footnote{
  Proposals for particle and nuclear physics experiments at J-PARC
  are available from \cite{JPARCproposals}.}.
The \KOTO  collaboration built the new neutral beam-line~\cite{KOTObeam}
(Fig.~\ref{Fig:KOTObeam})
at the Hadron Experimental Hall~\cite{HD-hall} of J-PARC
and surveyed the beam~\cite{KOTOshiomi,KOTOGOHmon} in 2009. 
They constructed a new electromagnetic calorimeter in 2010 
with the 27$X_0$-long undoped-CsI crystals~\footnote{
 The \KOTO calorimeter consists of 
 2240 small CsI blocks, in the size of 2.5$\times$2.5$\times$50 cm$^3$,
 for the central region 
 and 
 476 large CsI blocks, in the size of 5.0$\times$5.0$\times$50 cm$^3$, 
 for the outer region.
 }
,
used in the past in the KTeV experiment,
and started a commissioning~\footnote{
  J-PARC was affected by the East Japan Earthquake on March 11, 2011,
  but there was no effect of Tsunami that happened nearby and 
  no one was injured. The beam line and the calorimeter for \KOTO  was not damaged. 
  The J-PARC accelerators 
  started re-commissioning in December 2011 and resumed the operation in January 2012.
}. 
They will continue the detector construction in 2012
and expect to start the physics run in 2013.
The detector subsystems in E391a are reused or upgraded; 
in particular, 
the collar-shaped photon counter at the entrance of the decay region 
(CC02 in Fig.~\ref{Fig:E391adet} bottom) 
will be made of CsI crystals and renamed as Neutron Collar Counter, 
so as to be with the capability of measuring neutrons in the beam halo, 
and the counter to identify charged particles in front of the calorimeter 
(CV in Fig.~\ref{Fig:E391adet} bottom) 
will be made by plastic-scintillator strips.
The photon counter to cover the beam hole
(BA in Fig.~\ref{Fig:E391adet} bottom) 
will be replaced 
by an array of aerogel \v{C}erenkov counters with lead converters; 
the counters are designed to be insensitive to neutrons 
while keeping a high detection efficiency for photons.
The trigger and data acquisition systems  
are newly built
by introducing 
the schemes of waveform digitization 
and pipe-line readout.
As the first step in measuring $B(K^0_L \to \pi^0\nu\bar{\nu})$ at J-PARC, 
\KOTO aims at the first observation of the \KLpnn decay
at the SM sensitivity.

\begin{figure}[tb]
\begin{center}
\begin{minipage}[t]{16.5 cm}
\begin{center}
\epsfig{file=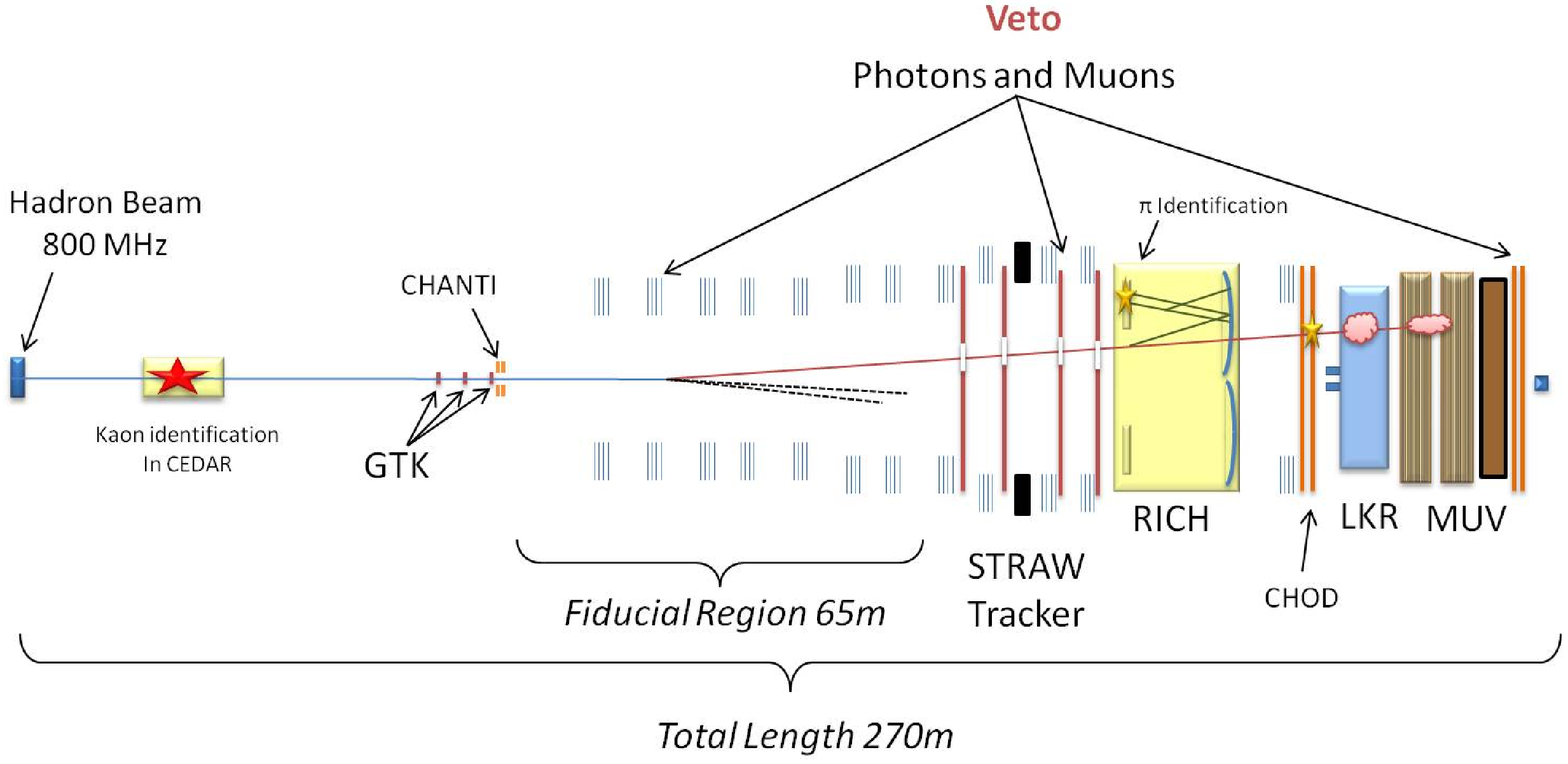,scale=0.60, angle=0}
\end{center}
\end{minipage}
\begin{minipage}[t]{16.5cm}
\caption{
Schematic view of the NA62 beam line and detector.
CEDAR: differential \v{C}erenkov counter;
GTK: silicon pixel tracking detectors;
CHANTI: guard-ring counters; 
RICH: Ring-Imaging \v{C}erenkov detector;
CHOD: charged-particle hodoscope;
LKR: liquid-Krypton calorimeter;
MUV: muon-veto detectors.
Source: Figure taken from Ref.\cite{NA62det}.
\label{Fig:NA62det}
}
\end{minipage}
\end{center}
\end{figure}

The next generation of \Kppnn
is the NA62 experiment~\cite{NA62web,NA62CKM2010} at CERN,
which will use $K^+$ decays in flight
from an un-separated beam of 75 GeV/$c$ from SPS.
The construction of the NA62 detector~\cite{NA62det,NA62detEPS2011} (Fig.~\ref{Fig:NA62det}) 
is proceeding steadily.
The incoming kaon is measured by the Gigatracker system in the beam.
The charged decay particle is measured by the straw-chamber spectrometer
and is identified by
the Ring Imaging \v{C}erenkov (RICH) detector and the muon-veto sampling calorimeter.
The LKr calorimeter, originally built for NA48,
is used 
as a veto for forward photons.
Photons at large angles are intercepted by a series of 12 ring-shaped veto counters 
constructed using lead-glass blocks from the OPAL electromagnetic barrel calorimeter.
A technical run will be made in the autumn of 2012
before the long shut-down of LHC.
Given CERN's current accelerator-operation plan,
first physics for NA62 is expected 
in 2014 or 2015.
The goal of NA62 is to detect 100 \Kppnn events
with no more than 10\% background.

In the US,
a new experiment to measure $B(K^+ \to \pi^+\nu\bar{\nu})$
with $K^+$ decays at rest
is proposed to FNAL
(P1021 ORKA).
Higher sensitivity kaon experiments based on a new high-intensity proton source
at FNAL~\cite{ProjectX,Tschirhart} are now under discussion~\cite{HighIntensity}.

\clearpage

\subsection{\it $K\to \ell\bar{\ell}$ and $K\to \pi \ell\bar{\ell}$
\label{subsec:Rare-Kpll}}

Rare kaon decays with charged leptons 
should be easier to detect in experiments 
because the kaon mass can be fully-reconstructed.
However, if a kaon decay accompanies 
charged leptons in the final state, 
the transition is also induced 
by long-distance effects with photon emission
in hadronic interactions (Section~\ref{sec:Radiative}); 
their theoretical interpretations are not straightforward.
The decays $K^0_L \to \mu^+ \mu^-$ and $K^0_L \to e^+ e^-$
are measured to be
$B(K^0_L \to \mu^+ \mu^-)$ = $(6.84\pm 0.11)\times 10^{-9}$
and 
$B(K^0_L \to e^+  e^-)$ = $(9^{+6}_{-4})\times 10^{-12}$, 
respectively~\cite{PDG2010update}; 
the latter is the smallest branching ratio yet measured in particle physics. 
However, 
these decay modes are saturated by an absorptive process~\cite{Sehgal69}:
$K^0_L \to \gamma\gamma$ and the two photons subsequently scattered into two leptons.
We basically look at a QED process, and cannot get good information
on the short distance contributions~\cite{mmIsidori}.

The KLOE experiment 
(to be discussed in Sections~\ref{subsec:Lepton-LFU} and \ref{sec:CPTQM})
obtained an upper limit of $9 \times 10^{-9}$~\cite{KLOEKsee}
for the branching ratio of the $K^0_S \to e^+ e^-$ decay.

To the $K^0_L \to \pi^0 \ell \bar{\ell}$  decay, 
there are four contributions~\footnote{
  Here we follow the discussions in \cite{peeBuchalla,pmmIsidori,MS11}
  and use the convention of  {\em direct} and {\em indirect} CP violation.
  As in $K^0_L\to\pi^0\nu\overline{\nu}$,
  {\em direct}  CP violation in $K^0_L \to \pi^0 \ell^+ \ell^-$
   corresponds to {\em interference between decays with and without mixing}. 
}
from direct CP violation,
indirect CP violation 
due to the $K_{1}$ component of $K^0_L$ (ICPV),
their interference (INT), 
and the CP conserving process (CPC)
through the $\pi^0\gamma^* \gamma^*$ intermediate state. 
The CPC contribution can be obtained from  the study of the
decay $K^0_L \to \pi^0 \gamma \gamma$~\cite{KLggNA48,KLggKTeVfinal}.
The  $K^0_S \to \pi^0 \ell \bar{\ell}$ decay, which is a CP conserving process of $K^0_S$, 
helps to do reliable estimation of ICPV (and INT)
and extract short-distance physics from 
$K^0_L \to \pi^0 \ell \bar{\ell}$
\cite{peeBuchalla,pmmIsidori}.
New Physics impacts on the $K^0_L \to \pi^0 \ell \bar{\ell}$  decays
are discussed in \cite{MS11,MST06}.

 The NA48/1 collaboration at CERN performed the data taking dedicated to 
 $K^0_S$ decays 
 with a high-intensity \KS beam: 
  $2\times 10^5$ \KS decays per spill with a mean energy of 120 GeV.
 The rare decays
 $K^0_S \to \pi^0 e^+ e^-$ and
 $K^0_S \to \pi^0 \mu^+ \mu^-$  were observed for the first time; 
 using a vector matrix element and unit form factor,
 the measured branching ratios were 
 $( 5.8^{+2.8}_{-2.3}(stat.) \pm 0.8(syst.) ) \times 10^{-9}$~\cite{NA48pi0ee}
 and 
 $( 2.9^{+1.5}_{-1.2}(stat.) \pm 0.2(syst.) )\times 10^{-9}$~\cite{NA48pi0mm}, 
 respectively.
 With these results and theoretical analyses, 
 $B(K^0_L \to \pi^0 e^+ e^-) = 3.23^{+0.91}_{-0.79} \times 10^{-11}\ $ 
      and
 $B(K^0_L \to \pi^0 \mu^+ \mu^-) = 1.29^{+0.24}_{-0.23} \times 10^{-11}\ $ 
  with $a_{S} > 0$
and 
 $B(K^0_L \to \pi^0 e^+ e^-) = 1.37^{+0.55}_{-0.43} \times 10^{-11}\ $ 
      and
 $B(K^0_L \to \pi^0 \mu^+ \mu^-) = 0.86^{+0.18}_{-0.17} \times 10^{-11}\ $ 
  with $a_{S} < 0$
 are predicted in the SM~\cite{MS11}, 
where $a_{S}$ is a parameter
of the form factors in the  $K^0_S \to \pi^0 \ell \bar{\ell}$ amplitude.

The $K^0_L \to \pi^0 e^+ e^-$  decay
has been studied by the KTeV experiment.
The limiting background was 
from the radiative Dalitz decay
$K^0_L \to e^+ e^- \gamma \gamma$  
($B = (5.95\pm 0.33)\times 10^{-7}$~\cite{PDG2010update}) 
with invariant mass of the two photons 
consistent with the nominal mass of $\pi^0$~\cite{Greenlee}.
Phase space cuts, which were applied to the data to suppress the background, 
reduced the signal acceptance by 25\%. 
The number of events observed in the signal region 
was consistent with the expected background 
for both of their 1997 and 1999 data sets. 
Combining these results, 
the KTeV final result was 
$B(K^0_L \to \pi^0 e^+ e^-) < 2.8\times 10^{-10}$~\cite{KTeVpee}.
KTeV also studied 
the $K^0_L \to \pi^0 \mu^+ \mu^-$ decay,
and has reported an upper limit 
$B(K^0_L \to \pi^0 \mu^+ \mu^-) < 3.8\times 10^{-10}$~\cite{KTeVpmm}
from the 1997 data set.
Both of the limits
are still an order of magnitude larger than the SM predictions.

The NA48/2 experiment measured the branching ratios of 
the $K^{\pm} \to \pi^{\pm} e^{+} e^{-}$
and $K^{\pm} \to \pi^{\pm} \mu^{+} \mu^{-}$ decays to be 
      $ ( 3.11 \pm 0.12 ) \times 10^{-7}$~\cite{NA48-piee}
and 
      $ ( 9.62 \pm 0.25 ) \times 10^{-8}$~\cite{NA48-pimm}, respectively.
The rates of these decays are dominated by the long-distance contributions
involving one photon exchange,
but the samples allow detailed studies of the decay properties
such as the charge asymmetries (Section~\ref{subsec:CP-decay}).

No new plan for the $K\to \ell\bar{\ell}$ and $K\to \pi \ell\bar{\ell}$ experiments
is being proposed.
The NA62 experiment is expected to collect samples of 
$K^{+} \to \pi^{+} e^{+} e^{-}$
and 
$K^{+} \to \pi^{+} \mu^{+} \mu^{-}$ decays
which will be significantly larger than the current world samples.

\clearpage

\section{Leptons in kaon decays
\label{sec:Lepton}}

This section covers the searches for explicit SM violations 
by measuring leptons in the final state of kaon decays.

\subsection{\it Transverse muon polarization
\label{subsec:Lepton-Trans}}

In the $K^+ \to \pi^0 \mu^+ \nu$  decay
 ($K^+_{\mu 3}$, 
  $B = (3.353\pm0.034)$\%~\cite{PDG2010update}), 
 the transverse muon polarization $p_t$
 (the perpendicular component of the muon spin vector relative to the decay plane
 determined by the momentum vectors of muon and pion 
 in the $K^+$ rest frame) is a T-odd quantity
 and is an observable of CP violation.
 Any spurious effect from final-state interactions is small
 ($<10^{-5}$),
 because no charged particle other than muon exists in the final state.
 $p_t$ is almost vanishing 
 ($\sim 10^{-7}$)~\cite{BigiSanda} in the SM,
 while new sources of CP violation 
 (e.g. due to interference between the charged-Higgs exchange and the W exchange amplitudes)
 may give rise to $p_t$
 as large as $10^{-3}$~\cite{GK1991,BigiSanda}.
 Thus the transverse muon polarization in  {$K^+_{\mu 3}$} has been regarded as
 a sensitive probe of non-SM CP violation,    
 and is a good example of looking beyond the SM
 by measuring a decay property with high statistics.

\begin{figure}[tb]
\begin{center}
\begin{minipage}[t]{16.5cm}
\begin{center}
\epsfig{file=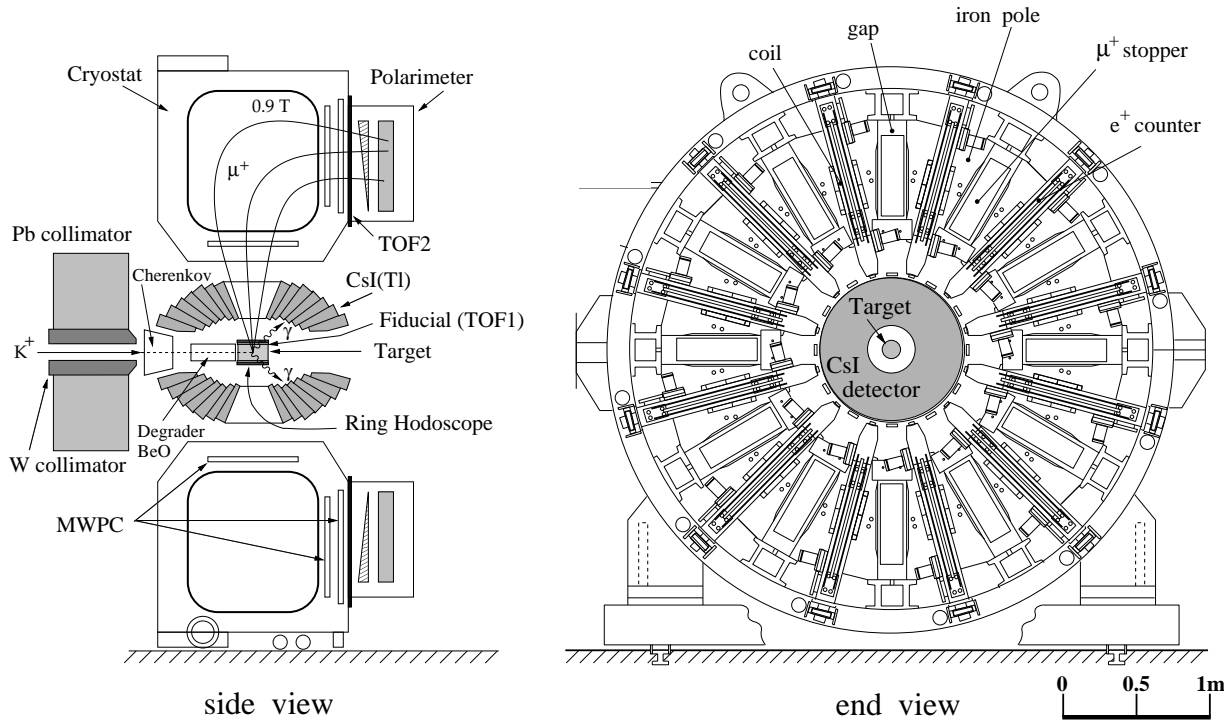,scale=0.60}
\end{center}
\end{minipage}
\begin{minipage}[t]{16.5cm}
\caption{
General assembly side and end views of the E246 detector.
MWPC: multiwire proportional chamber for charged-particle tracking;
TOF: time-of-flight counters for particle identification;
CsI(Tl): thallium-doped CsI crystals to detect two photons from $\pi^0$ decay.
Source: Figures taken from Ref.\cite{E246det}.
\label{Fig:E246det}
}
\end{minipage}
\end{center}
\end{figure} 

 The E246 collaboration at KEK
 measured the charged track and photons
 from $K^+$ decays at rest.
 The E246 detector~\cite{E246det} is shown in Fig.~\ref{Fig:E246det}. 
 They used the superconducting toroidal spectrometer 
 (consisting of 12 identical spectrometers arranged in 
 rotational symmetry),
 which enabled the experiment 
 to control possible sources of systematic uncertainties
 in polarization measurement. 
 They accumulated 11.8M events and obtained 
 $p_t=(-0.17\pm 0.23(stat.) \pm 0.11 (syst.) ) \times 10^{-2}$~\cite{E246final} 
 by the total data sets of E246 from 1996 to 2000, 
 giving an upper limit $|p_t|<0.0050$.
They also performed the first measurement of $p_t$ 
in the decay $K^+\to\mu^+\nu\gamma$: 
 $p_t=(-0.64\pm 1.85 (stat.) \pm 0.10 (syst.) ) \times 10^{-2}$~\cite{E246km2g}.

A successor to E246
is another new kaon experiment at J-PARC, E06 TREK~\cite{TREKweb,TREKimazato}, 
aiming at a $p_t$ sensitivity of $10^{-4}$.
The E246 superconducting toroidal magnet will be used,
and the detector will be upgraded 
by introducing the  Gas Electron Multiplier (GEM) tracker and the active muon-polarimeter.
A low-momentum beam line named K1.1BR
was built at the Hadron Experimental Hall
and was successfully commissioned in 2010.

\subsection{\it Lepton flavor violation
\label{subsec:Lepton-LFV}}

Experimental search for lepton-flavor violating (LFV) kaon decays
$K \to \mu e$ and $K\to \pi \mu e$
has a long history. 
The kaon system is well suited to the investigation of 
LFV new processes
involving both quarks and charged leptons~\footnote{
   Assuming an additive quantum number for quarks and leptons in the same generation
    (``one'' for down-quark and electron, ``two'' for strange-quark and muon, ...),
   the net number is conserved in the LFV kaon decays.}
due to the high sensitivity achieved by experiments on this system.
The mass of a hypothetical gauge boson for the tree-level effects 
should be in the scale of a few hundred 
TeV/$c^2$~\cite{LFVtheory1,LFVtheory2}.
A drawback is that
the LFV processes induced by Supersymmetric loop effects 
are not so promising, 
because ``Super-GIM'' suppression mechanism is 
expected in both quark and lepton sectors~\cite{LFVSUSY}.

Both two-body and three-body decays have to be explored 
in spite of the phase-space difference, 
because the $K \to \pi \mu e$ decay is 
sensitive to vector and scalar interactions. 
The upper limits on 
the $K^0_L \to \mu^{\pm} e^{\mp}$
and 
$K^+ \to \pi^+ \mu^+ e^-$ branching ratios
are 
$4.7\times 10^{-12}$
and $1.3\times 10^{-11}$,
respectively~\footnote{
      The limit $4.7 \times 10^{-12}$ 
      on $B(K^0_L \to \mu^{\pm} e^{\mp})$ 
      had been the most stringent upper limit to particle decays,
      until the MEG collaboration recently published an upper limit of $2.4\times 10^{-12}$
      on the $\mu^+\to e^+ \gamma$ branching ratio~\cite{MEG}. 
 }
~\cite{PDG2010update,E865kpmn}.
The KTeV experiment set the upper limits for the branching ratios 
$B(K^0_L\to \pi^0 \mu^{\pm} e^{\mp}) < 7.6 \times 10^{-11}$ 
and 
$B(K^0_L\to \pi^0 \pi^0 \mu^{\pm} e^{\mp}) < 1.7 \times 10^{-10}$
in \cite{KTeVLFV}.

No new plan for the search for LFV kaon decays
is being proposed.

\subsection{\it Lepton flavor universality
\label{subsec:Lepton-LFU}}

\begin{figure}[tb]
\begin{center}
\begin{minipage}[t]{16.5cm}
\begin{center}
\epsfig{file=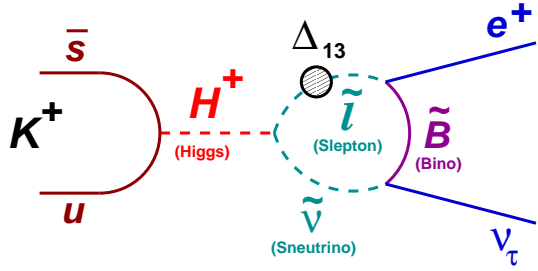,scale=0.40}
\end{center}
\end{minipage}
\begin{minipage}[t]{16.5cm}
\caption{
Diagram for the $K^+\to e^+ \nu_\tau$ decay
due to a new process induced by 
a charged-Higgs particle 
and an LFV loop of Supersymmetric particles.
$\Delta_{13}$ represents the effective $e$-$\tau$ coupling constant
in the loop.
Source: Figure by courtesy of the NA62 collaboration.
\label{Fig:LFUdiagram}
}
\end{minipage}
\end{center}
\end{figure} 

The LFV kaon decays are currently studied, intensively,
in the context of high precision tests of 
lepton flavor universality in the purely leptonic charged-kaon decay. 
The ratio $R_{K}
\equiv \Gamma(K^+\to\ e^+\nu(\gamma)) / \Gamma(K^+\to\ \mu^+\nu(\gamma))$
is helicity suppressed in the SM
due to the V-A couplings
and is predicted to be 
$R_{K}^{SM} = (2.477\pm 0.001)\times 10^{-5}$~\cite{CiriglianoRosell}, 
in which the radiative decay $K^+\to e^+ \nu\gamma$ ($K_{e2\gamma}$) via
internal bremsstrahlung  is included. 
On the other hand,
suppose 
an LFV new decay $K^+\to e^+ \nu_{\tau}$ exists due to the process of 
an intermediate charged-Higgs particle 
and an LFV Supersymmetric loop
(Fig.~\ref{Fig:LFUdiagram})~\cite{MPP06}.
Since the neutrino flavor is undetermined experimentally,
the measured $R_{K}$ should be regarded as
\be
R_{K}\ =\ 
\frac{\Sigma_i \ \Gamma(K^+\to\ e^+\nu_i)}
        {\Sigma_i \ \Gamma(K^+\to\ \mu^+\nu_i)}
\ \simeq\ 
  \frac{\Gamma_{SM}(K^+\to\ e^+\nu_{e}) + \Gamma_{NP}(K^+\to\ e^+\nu_{\tau}) } 
          {\Gamma_{SM}(K^+\to\ \mu^+\nu_{\mu})}
\label{eq:LFUratio}
\ee
assuming that 
$\Gamma_{NP}(K^+\to\ e^+\nu_{\mu})$ is small.
The deviations from the SM prediction,
denoted as $\Delta R_{K}$, 
in the relative size of $10^{-2} \sim 10^{-3}$ are suggested~\cite{MPP08}.
In such a case, 
with the charged Higgs mass $m_{H^{+}}$, 
the ratio of the Higgs vacuum expectation values for the up- and down- quark masses
(denoted as $\tan \beta$), 
and the effective $e-\tau$ coupling constant $\Delta_{13}$, 
the size of $\Delta R_{K}$ is predicted as
\be
   \frac{\Delta R_{K}}{R_{K}^{SM}}\ =\
      \frac{\Gamma_{NP}(K^+\to\ e^+\nu_{\tau})}{\Gamma_{SM}(K^+\to\ e^+\nu_{e})}\ =\ 
   \left( \frac{m_{K^{+}}}{m_{H^{+}}}   \right)^4\ 
   \left( \frac{m_{\tau}}{m_{e}} \right)^2\
   | \Delta_{13} |^2\
   \tan^6 \beta 
\label{eq:LFUtanbeta}
\ee
and can be experimentally studied.
The lepton flavor universality is also pursued 
by the PIENU experiment~\cite{PIENU} at TRIUMF
and the PEN experiment~\cite{PEN} at PSI
to measure 
$\Gamma(\pi^+\to\ e^+\nu(\gamma)) / \Gamma(\pi^+\to\ \mu^+\nu(\gamma))$
precisely.

\begin{figure}[tb]
\begin{center}
\begin{minipage}[t]{16.5cm}
\begin{center}
\epsfig{file=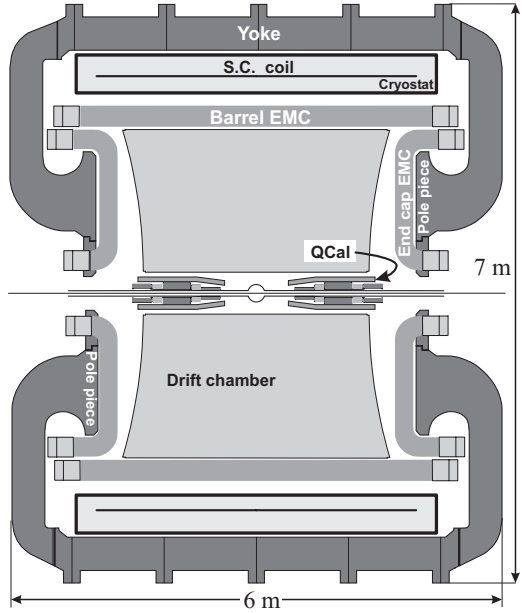,scale=0.40}
\end{center}
\end{minipage}
\begin{minipage}[t]{16.5cm}
\caption{
Vertical cross-section of the KLOE detector.
QCal: lead/scintillator-tile calorimeters to detect photons
that would be otherwise absorbed on the permanent quadrupole magnets for beam focusing;  
EMC: electromagnetic calorimeter; 
S.C. coil: superconducting coil.
Source: Figure taken from Ref.\cite{KLOEreview3}.
\label{Fig:KLOEdet}
}
\end{minipage}
\end{center}
\end{figure} 

\begin{figure}[tb]
\begin{center}
\begin{minipage}[t]{16.5cm}
\begin{center}
\epsfig{file=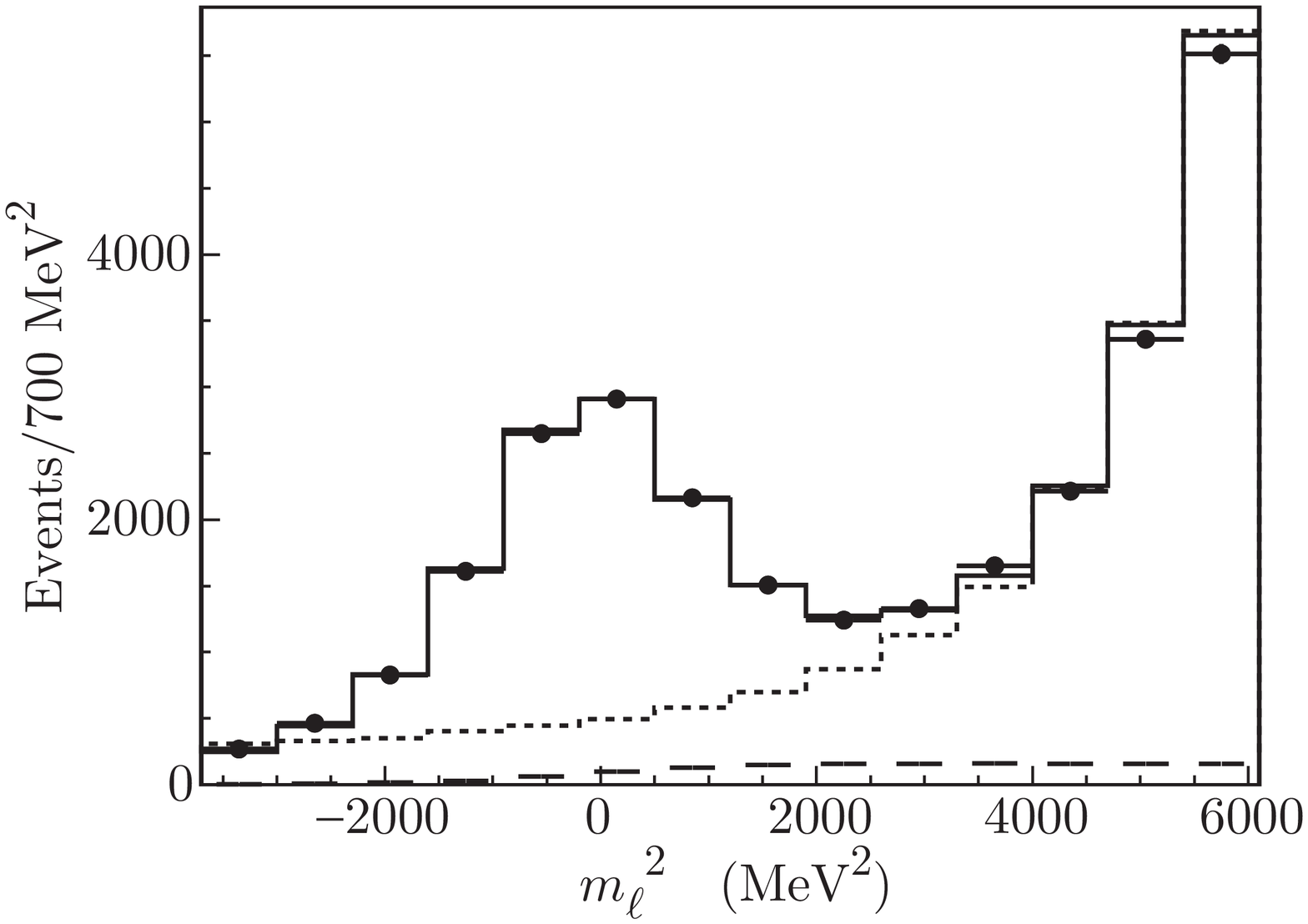,scale=0.35}
\hspace*{1.0cm}
\epsfig{file=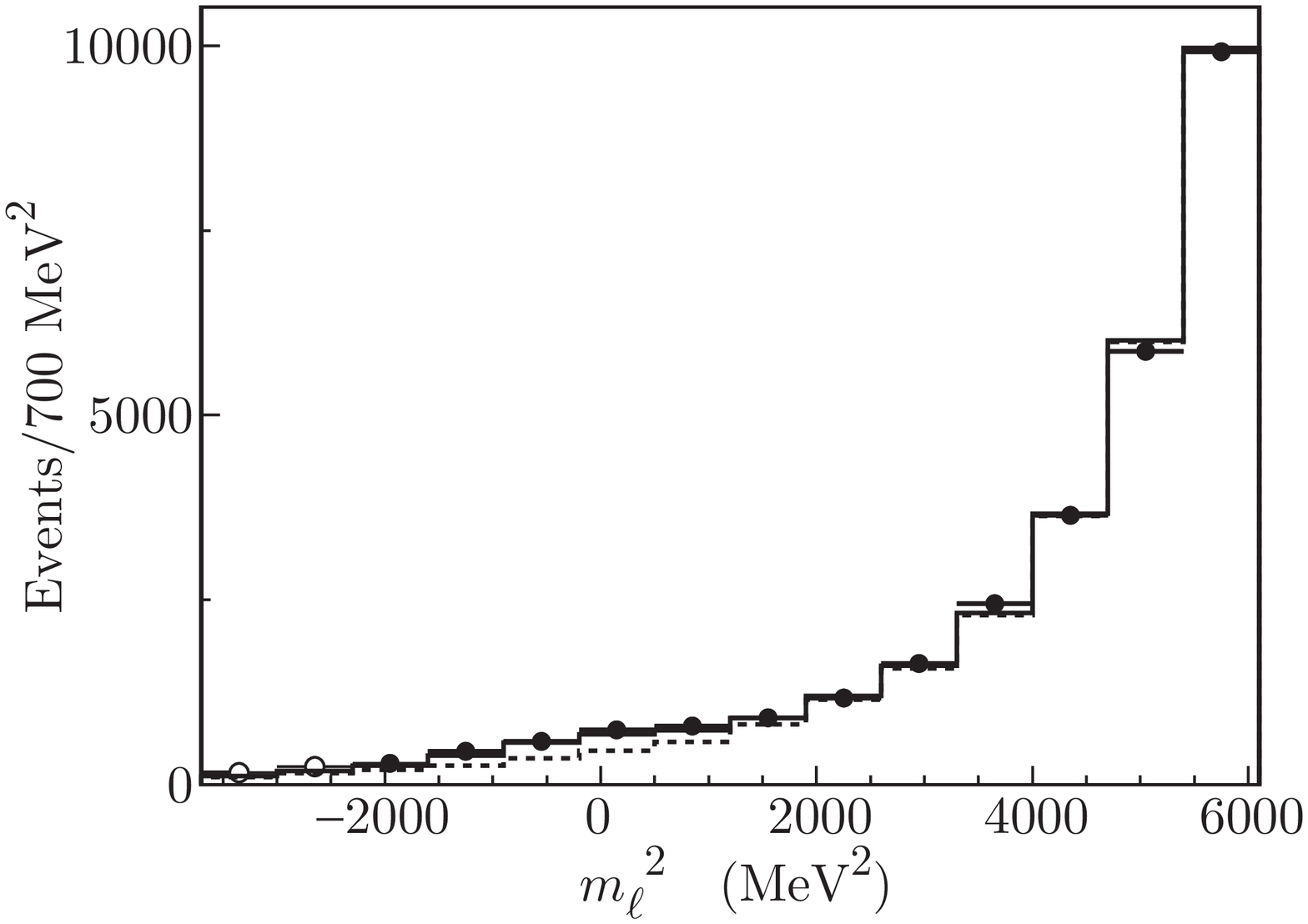,scale=0.35}
\vspace*{0.15cm}
\end{center}
\end{minipage}
\begin{minipage}[t]{16.5cm}
\caption{
KLOE's $K^{\pm} \to\ e^{\pm}\nu$ analysis
with the neural net output $NN$: 
the sum of fit results for $K^+$ and $K^-$ projected 
onto the $m_{\ell}^2$ axis in a signal region ($NN > 0.98$, left) 
and a background region ($0.86 < NN < 0.98$, right),
for data (black dots), Monte Carlo fit (solid line), and
$K_{\mu 2}$ background (dotted line).
The contribution from $K_{e 2}$ events with $E_{\gamma}>10$ MeV
is shown in dashed line.
Source: Figures taken from Ref.\cite{KLOE-RK}.
\label{Fig:KLOEke2}
}
\end{minipage}
\end{center}
\end{figure} 

The KLOE collaboration at DA$\Phi$NE, 
the Frascati $\phi$ factory~\cite{KLOEreview1,KLOEreview2,KLOEreview3},
measured $R_{K}$ 
with a data set consisting of 2.2 fb$^{-1}$
collected during 2001-2005~\cite{KLOE-RK},
corresponding to 3.3G $K^+K^-$ pairs produced from $\phi$ meson decays.
\DAFNE is an $e^+ e^-$ collider operated at 
a total energy $\sqrt{s} = m_{\phi}\cdot c^2 \sim $1.02 GeV.
$\phi$ mesons were produced with a cross section of $\sim 3.1$ $\mu$b
and decayed into $K^+K^-$ pairs and $K^0_L K^0_S$ pairs 
with branching ratios of 49\% and 34\%, respectively;
thus, \DAFNE is a copious source of {\em tagged} and {\em monochromatic} kaons. 
The KLOE detector (Fig.~\ref{Fig:KLOEdet}),
whose radius was 3.5 m to catch 63\% of all decaying $K^0_L$'s,
consisted of the cylindrical drift chamber and electromagnetic calorimeters
with an axial magnetic field of 0.52 T.
The $K^{\pm} \to\ \ell^{\pm}\nu$ decay in flight ($\sim$0.1 GeV/$c$) 
was reconstructed 
by the tracks of a kaon and a decay product, with the same charge,  
in the drift chamber, and
the squared mass $m_{\ell}^2$
of the lepton for the decay :
\be
  m_{\ell}^2\ =\ (\ E_K\ -\ | \vec{p}_K - \vec{p}_d |\ )^2\ -\ (\ \vec{p}_d\ )^2
\label{eq:KLOEke2}
\ee
was computed
from the measured kaon and decay-particle momenta $\vec{p}_K$ and $\vec{p}_d$ .
In the analysis,
the $K^{\pm} \to\ e^{\pm}\nu$ events at around $m_{\ell}^2=0$
should be distinguished 
from the background contamination
due to the tail of the prominent $K^{\pm} \to\ \mu^{\pm}\nu$ peak; 
the kinematic and track quality cuts were imposed 
and then
the information about shower profile and total energy deposition 
in the calorimeter, combined with a neural network technique,
and the time-of-flight information were used 
for electron identification (Fig.~\ref{Fig:KLOEke2}).
The numbers of $K\to e\nu(\gamma)$ events were 
$7064\pm 102$ for $K^+$ and
$6750\pm 101$ for $K^-$, respectively, 
89.8\% of which had $E_{\gamma}<10$ MeV
and contained both 
the $K\to e\nu$ events and the $K_{e2\gamma}$ events with $E_{\gamma}<10$ MeV.
No photon detection requirement was imposed; 
the contribution from the $K_{e2\gamma}$ events with $E_{\gamma}>10$ MeV, 
due to the direct-emission process, 
was studied~\cite{KLOE-RK,KLOE-rad} by using a separate sample 
with photon detection requirement, 
and was subtracted.
The numbers of $K\to \mu\nu(\gamma)$ events were 
287.8M for $K^+$ and
274.2M for $K^-$, respectively.
The difference between the $K^+$ and $K^-$ counts was
due to the larger cross section of $K^{-}$ nuclear interaction 
in the material traversed.
Finally, KLOE obtained~\footnote{
  The rate $R_{10}$,
  defined as $R_{10}\ =\ \Gamma(K\to e \nu (\gamma), E_{\gamma}< 10~{\rm MeV}) / \Gamma (K\to \mu \nu)$, 
  was measured and was converted into $R_{K}$ with the relation  
  $R_{10} = R_{K} \times (0.9357 \pm 0.0007)$.
} 
$R_{K}\ =\  ( 2.493 \pm 0.025(stat.) \pm 0.019 (syst.) ) \times 10^{-5}$,
in agreement with the SM prediction.

\begin{figure}[tb]
\begin{center}
\begin{minipage}[t]{16.5cm}
\begin{center}
\epsfig{file=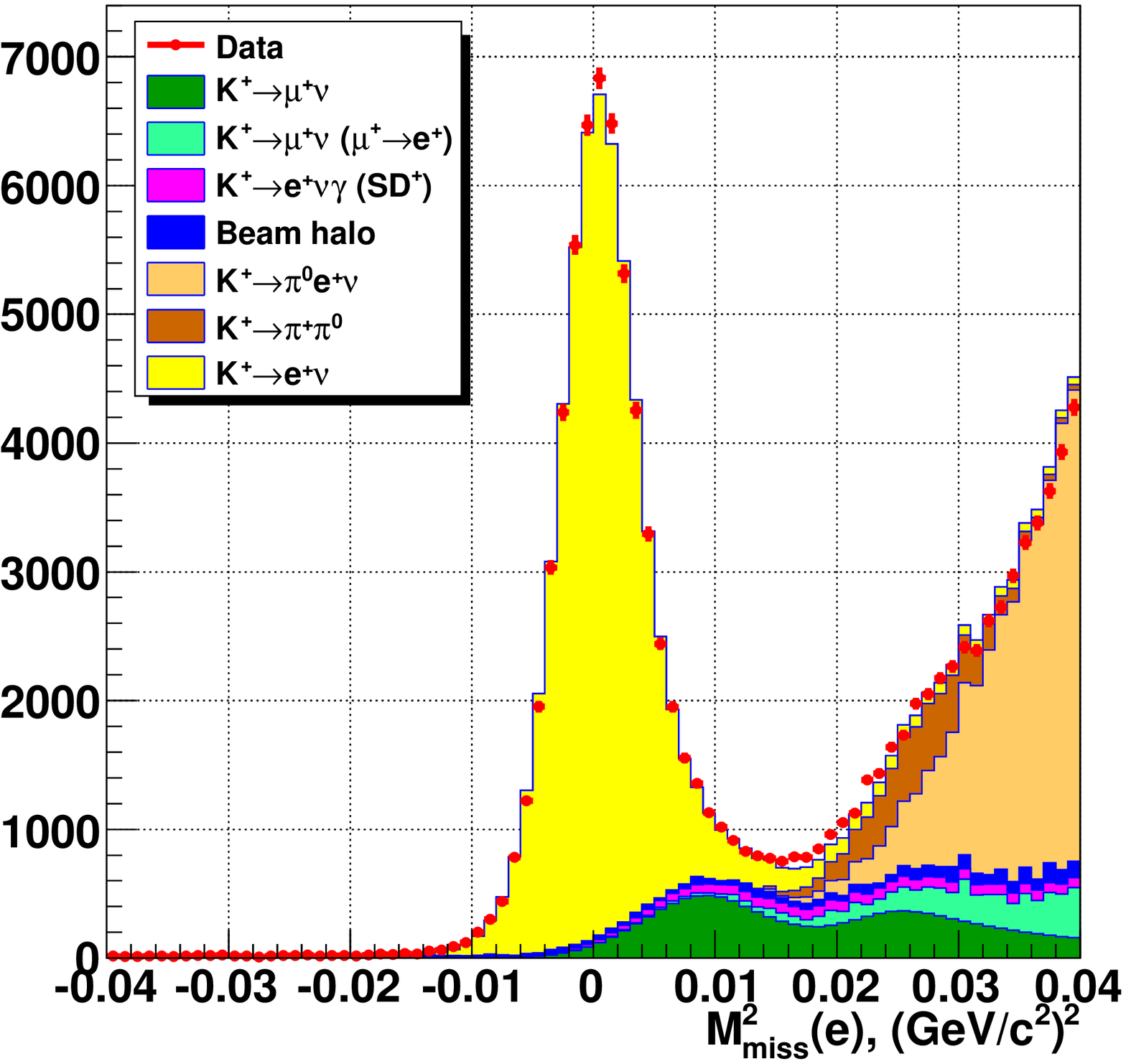,scale=0.35}
\hspace*{1.0cm}
\epsfig{file=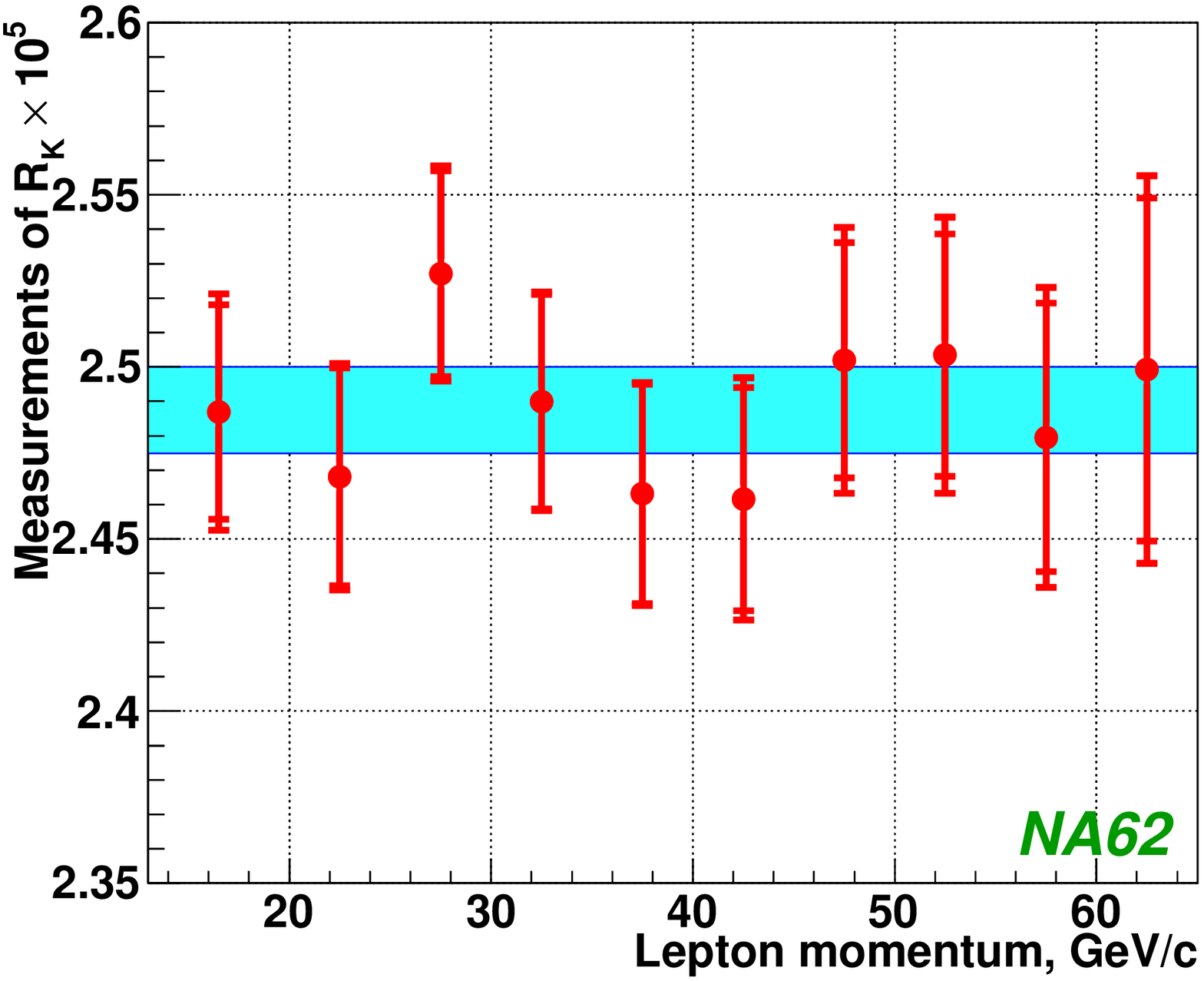,scale=0.35}
\vspace*{0.15cm}
\end{center}
\end{minipage}
\begin{minipage}[t]{16.5cm}
\caption{
NA62's $K^{\pm} \to\ e^{\pm}\nu$ analysis:
reconstructed squared missing-mass $M_{miss}^2(e)$
distribution of the $K^+\to\ e^+\nu$ candidates 
compared with the sum of normalized estimated signal and background components (left);
measurements of $R_K$ in lepton momentum bins,
with the average $R_K$ and its total uncertainty indicated by a band (right).
Source: Figures taken from Ref.~\cite{NA62-RK}.
\label{Fig:NA62ke2}
}
\end{minipage}
\end{center}
\end{figure} 

The NA62 collaboration at CERN, 
in the first phase,
collected data to measure $R_{K}$ during four months in 2007, 
and collected special data samples to study systematic effects for two weeks in 2008.
The beam line and setup of NA48/2 were used; 
90\% of the data were taken with the un-separated $K^+$ beam 
of $74.0\pm 1.4{\rm (rms)}$ GeV/$c$,
because the muon sweeping system optimized for the positive beam
provided better suppression 
of the positrons produced by beam halo muons via $\mu\to e$ decay.
The LKr calorimeter was used for lepton identification 
as well as a photon veto with the energy threshold of 2 GeV.
Results based on the analysis of 40\% of the 2007 data 
collected with the $K^+$ beam only
were published in~\cite{NA62-RK}.
The kinematic identification of $K^+\to \ell^+\nu$ decays is 
based on the reconstructed squared missing-mass $M_{miss}^2(\ell)$
in the positron ($\ell = e$) or muon ($\ell = \mu$) hypothesis: 
\be
  M_{miss}^2\ =\ (\ P_K\ -\ P_{\ell}\ )^2
\label{eq:NA48ke2}
\ee
from the average kaon four-momentum $P_K$
(monitored in time with $K^+\to\pi^+\pi^+\pi^-$ decays)
and the reconstructed lepton four-momentum $P_{\ell}$
under the positron or muon mass hypothesis (Fig.~\ref{Fig:NA62ke2} left).
The number of reconstructed $K^+\to e^+\nu$ candidate events was $59813$,
and the number of $K^+\to \mu^+\nu$ candidate events was
18.03M.
The total background contamination was $(8.71\pm 0.24)$ \%; 
the main source of the  background
was found to be the $K^+ \to \mu^+ \nu$ decay 
with muon identification as positron
due to {\em catastrophic} bremsstrahlung
in or in front of the LKr calorimeter.
The probability of the mis-identification
was studied with pure muon samples, 
without positron contamination (typically $\sim 10^{-4}$)
due to $\mu^+\to e^+$ decays in flight. 
The sample was selected from the tracks traversing a $9.2X_0$-thick lead wall,  
covering 20\% of the geometrical acceptance,  
which was installed 1.2 m in front of the LKr calorimeter
during a dedicated period. 
The $K^+ \to \mu^+ \nu$ background contamination
was estimated to be 
$(6.11\pm 0.22)$ \%.
Due to the significant dependence of acceptance and background on lepton momentum, 
the $R_{K}$ measurement was performed independently in ten momentum bin
covering a range from 13 to 65 GeV/$c$ (Fig.~\ref{Fig:NA62ke2} right).
The selection criteria had been optimized separately in each momentum bin.
$R_{K}$ was obtained to be 
$( 2.487 \pm 0.011(stat.) \pm 0.007 (syst.) ) \times 10^{-5}$,
consistent with the SM  prediction.

\begin{figure}[tb]
\begin{center}
\begin{minipage}[t]{16.5cm}
\begin{center}
\epsfig{file=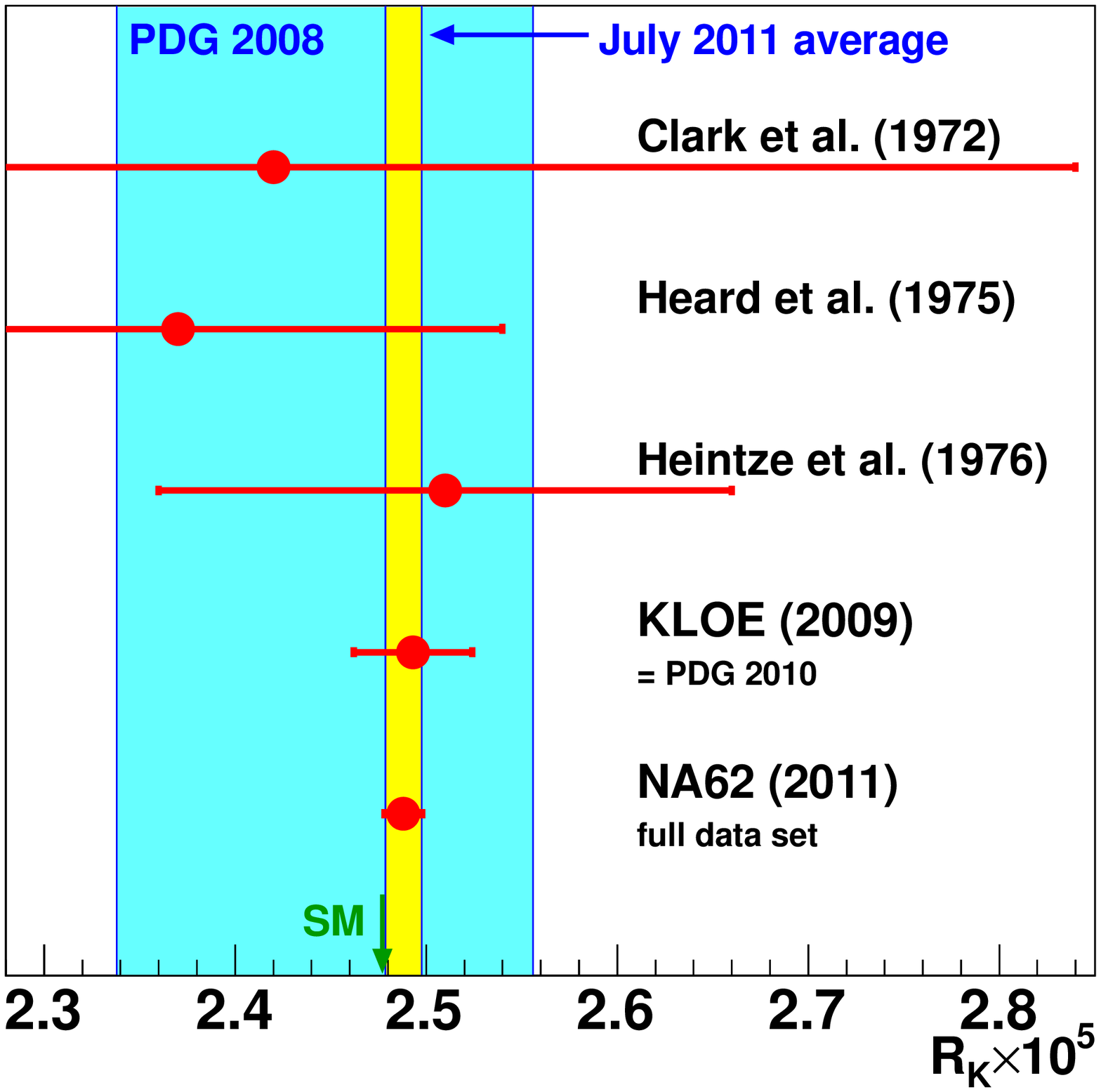,scale=0.35}
\hspace*{1.0cm}
\epsfig{file=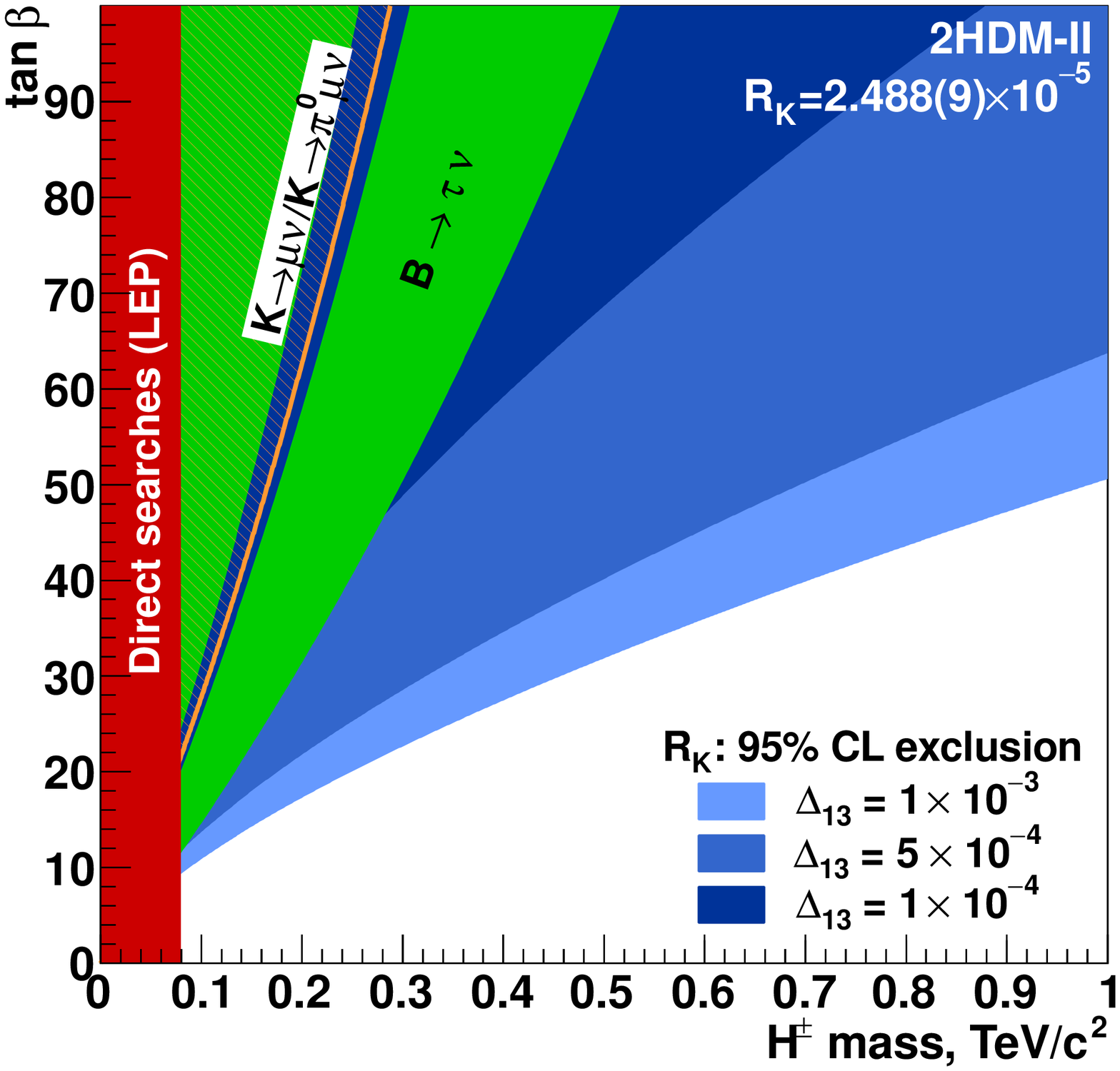,scale=0.35}
\vspace*{0.15cm}
\end{center}
\end{minipage}
\begin{minipage}[t]{16.5cm}
\caption{
Summary of the $R_K$ measurements 
including NA62's new results in \cite{NA62Ke2EPS2011} (left); 
excluded regions at 95\% C.L. in the ($m_{H^{+}}$,$\tan \beta$) plane
for $\Delta_{13}$=$10^{-4}$, $5\times 10^{-4}$ and $10^{-3}$
with the constraints from other experiments (right).
Source: Figures by courtesy of the NA62 collaboration.
\label{Fig:LFUresults}
}
\end{minipage}
\end{center}
\end{figure} 

The NA62 collaboration recently reported 
results based on 
the analysis of the full data set~\cite{NA62Ke2EPS2011},
superseding the results in~\cite{NA62-RK}.
They obtained 
$R_{K} = 
( 2.488 \pm 0.007(stat.) \pm 0.007 (syst.) ) \times 10^{-5}$ and,
combining with other $R_K$ measurements, 
the world average is
$(2.488\pm 0.009) \times 10^{-5}$, with 0.4\% precision,
as presented in Fig.~\ref{Fig:LFUresults} left.
The regions excluded at 95\% C.L. in the ($M_{H^{+}}$,$\tan \beta$) plane
are shown in Fig.~\ref{Fig:LFUresults} right, 
for different values of $\Delta_{13}$.

The $R_K$ measurement will continue 
in the next phase of NA62 for $K^+\to\pi^+\nu\bar{\nu}$
as well as in the future 
KLOE-2 experiment (Section~\ref{sec:CPTQM}).
The TREK collaboration at J-PARC has also submitted a proposal P36~\cite{TREKP36},
as their first phase, 
to measure $R_{K}$
with $K^+$ decays at rest.

\clearpage

\section{Tests of CPT and quantum mechanics
\label{sec:CPTQM}}

In the $\phi$ decay into  $K^0_L K^0_S$ pairs, 
the initial state is a coherent (and entangled) quantum state:
\be
 |i\rangle\ 
      =\  \frac{1}{\sqrt{2}}\ [\  |K^0\rangle |\overline{K^0}\rangle\ -\ |\overline{K^0}\rangle |K^0\rangle \ ]\  
      =\  \frac{N}{\sqrt{2}}\ [\  |K^0_S\rangle |K^0_L\rangle\ -\ |K^0_L\rangle |K^0_S\rangle \ ]
\label{eq:KLentangle}
\ee
where $N\simeq 1$ is a normalization factor. 
In the KLOE experiment, 
by tagging the events with $K^0_L$ reaching the calorimeter without decaying
(dubbed ``K-{\em crash}"), 
a pure $K^0_S$ beam was available and 
there has been a major improvement in $K^0_S$ decay measurements
(to be discussed in Section~\ref{sec:Basic}).
With the results of various new measurements on neutral kaon decays,
the Bell-Steinberger relation was used~\cite{KLOE-BS} 
to provide a constraint 
relating the unitarity of the sum of the decay amplitudes to the CPT observables.
The latest limit on the mass difference between $K^0$ and $\overline{K^0}$ 
was $4.0\times 10^{-19}$ GeV at the 95\% C.L.~\cite{KLOE-CPT}.

\begin{figure}[tb]
\begin{center}
\begin{minipage}[t]{16.5cm}
\begin{center}
\epsfig{file=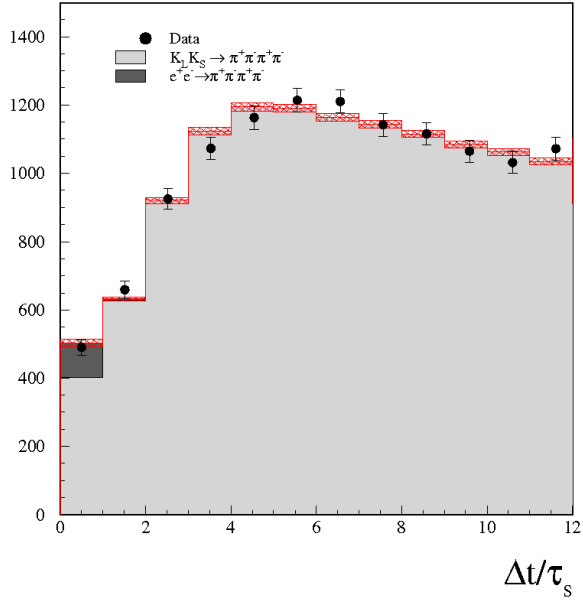,scale=0.40}
\end{center}
\end{minipage}
\begin{minipage}[t]{16.5cm}
\caption{
KLOE result on the fit to the measured $I(\Delta t)$ distribution 
               of $\phi \to K^0_S K^0_L\to \pi^+\pi^-\pi^+\pi^-$.
               The black points with errors are data and the solid histogram is the fit result. 
               The uncertainty arising from the efficiency correction is shown as the hatched area.
Source: Figure taken from Ref.\cite{KLOE-dec}.
\label{Fig:QMdist}
}
\end{minipage}
\end{center}
\end{figure} 

In the CP-violating process 
$\phi \to K^0_S K^0_L\to \pi^+\pi^-\pi^+\pi^-$, 
KLOE observed the quantum interference between two kaons for the first time~\cite{KLOE-int}. 
The measured $\Delta t$ distribution,
with $\Delta t$ the absolute value of the time difference of the two $\pi^+\pi^-$ decays, 
can be fitted with the distribution in the $K^0_S-K^0_L$ basis: 
\bea
I(\Delta t) & \propto & 
      e^{- \Gamma_L \Delta t}\ +\ e^{- \Gamma_S \Delta t}\ 
        -\ 2(1-\zeta_{SL})\ e^{-  \frac{(\Gamma_S+\Gamma_L)}{2} \Delta t}\ \cos (\Delta m \Delta t)   \\
                 & \rightarrow &
            2 \zeta_{SL} \ \left( 1-  \frac{(\Gamma_S+\Gamma_L)}{2} \Delta t \right)
                      \ \ \ \ \ \ \ \ \ \ \Delta t\rightarrow 0
\label{eq:QMtest}
\eea
where $\Delta m$ is the mass difference between $K^0_L$ and $K^0_S$.
The interference term 
$e^{-  \frac{(\Gamma_S+\Gamma_L)}{2} \Delta t}\ \cos (\Delta m \Delta t)$
is multiplied by a factor $(1-\zeta_{SL})$
with a {\em decoherence} parameter $\zeta_{SL}$, which represents  
a loss of coherence during the time evolution of the states
and should be zero in quantum mechanics (QM).
Final results obtained from KLOE
with 1.5 fb$^{-1}$ in 2004-2005 were~\cite{KLOE-dec}
$\zeta_{SL}\ =\ ( 0.3 \pm 1.8(stat.) \pm 0.6 (syst.) ) \times 10^{-2}$
(Fig.~\ref{Fig:QMdist}) and,
to the fit with the distribution in the $K^0-\overline{K^0}$ basis,
$\zeta_{0\overline{0}}\ =\ ( 1.4 \pm 9.5(stat.) \pm 3.8 (syst.) ) \times 10^{-7}$;
no deviation from QM was observed~\footnote{
   Decoherence in the $K^0-\overline{K^0}$ basis
   results in the CP-allowed 
   $K^0_S K^0_S\to \pi^+\pi^-\pi^+\pi^-$ decays and thus
   the value for the decoherence parameter $\zeta_{0\overline{0}}$ is much smaller
    than $\zeta_{SL}$.
   }.
Since the measurement of non-zero $\zeta_{SL}$ is sensitive to the distribution
in the small $\Delta t$ region, 
the decay vertex resolution due to charged-track extrapolation
($\simeq 6$mm in the KLOE detector, corresponding to $\sim 1\tau_{S}$, 
 to the vertex of a $K^0_S\to\pi^+\pi^-$ decay close to the interaction point)
should be improved
in future experiments. 
Other tests of CPT invariance and the basic principles of QM 
are discussed in~\cite{KLOE-dec,CPTQM}.

Kaon physics at the $\phi$ factory will continue 
with an upgraded KLOE detector, KLOE-2~\cite{KLOE-2web,KLOE-2doc}.
In 2008, 
\DAFNE was upgraded with a new interaction scheme 
({\em Crabbed Waist} collisions~\cite{CrabbedWaist2007}).
During 2008-2009 a factor-of-three larger peak-luminosity 
($4.53\times 10^{32}$ cm$^{-2}$s$^{-1}$)
and a factor-of-two larger integrated luminosity per an hour (1 pb$^{-1}$) 
than what previously obtained
were achieved~\cite{Zobov2010}.
The \DAFNE commissioning is scheduled to be resumed in October 2011.
In the first phase,
the detector with minimal upgrade
(two devices along the beam line to tag the scattered electrons/positrons 
from $\gamma\gamma$ collisions) 
restarts taking data. 
To the next phase, 
a cylindrical GEM detector~\cite{KLOE2-GEM}
will be placed between the beam pipe and the inner wall of the drift chamber,  
as a new Inner Tracker, 
to improve the decay vertex resolution
and to increase the acceptance for low transverse-momentum tracks.
A crystal calorimeter with timing (CCALT) for low-angle and low-energy photons
and a quadrupole calorimeter with tiles (QCALT) 
to cover the area of the final focusing section
will also be installed.
From 2013 to 2015, 
physics run to integrate 20 fb$^{-1}$ with the KLOE-2 detector 
is expected~\cite{KLOE-2EPS2011phys,KLOE-2EPS2011det}.

\clearpage

\section{Radiative decays
\label{sec:Radiative}}

\begin{figure}[tb]
\begin{center}
\begin{minipage}[t]{16.5cm}
\begin{center}
\epsfig{file=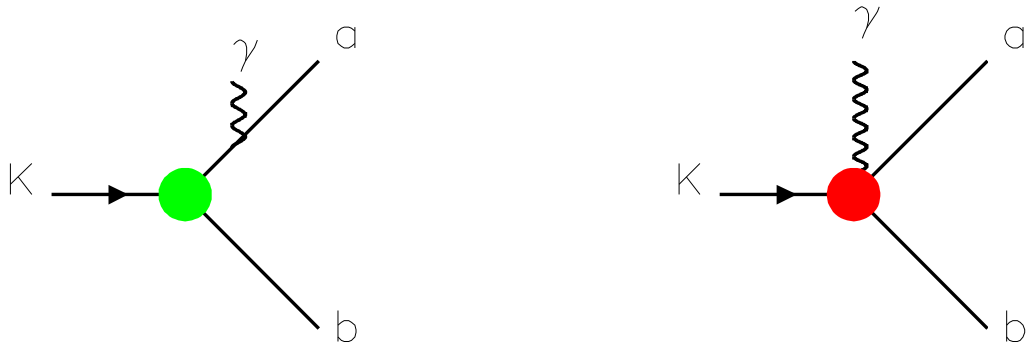,scale=0.85}
\end{center}
\end{minipage}
\begin{minipage}[t]{16.5cm}
\caption{
Graphs of the IB (left) and SD or DE (right) components of the radiative $K\to ab$ decay.
\label{Fig:RADgraph}
}
\end{minipage}
\end{center}
\end{figure} 

A {\em radiative} kaon decay,
which is accompanied by real photons in the final state, 
is due to radiative effects in the leptonic, semileptonic, or non-leptonic transitions.
The decay can proceed via 
\begin{itemize}
 \item inner bremsstrahlung (IB),
           in which a photon is emitted from the charged particle in the initial or final state
           (Fig.~\ref{Fig:RADgraph} left), 
 \item structure-dependent (SD) or direct-emission (DE) radiative decay~\footnote{
          Traditionally, 
          the emission in the leptonic or semileptonic transition is called structure dependent
          (because it is sensitive to the hadronic structure of kaon) 
          and the emission in the non-leptonic transition is called direct emission
          (because it is due to hadronic interactions of mesons); 
          however, not all the papers follow the convention.
 },
   which involves the emission of a photon
   from the intermediate states in the transition (Fig.~\ref{Fig:RADgraph} right), and 
  \item their possible interference (INT). 
\end{itemize}
Processes where a photon internally converts 
into an  $e^+ e^-$ pair or a $\mu^+ \mu^-$ pair (if kinematically allowed),
called a Dalitz pair~\cite{Dalitz}, 
can also be considered; 
the plane of the Dalitz pair in the kaon rest frame
contains information on the polarization of the virtual photon. 

Radiative kaon decays occur at the branching ratio of $10^{-3}\sim 10^{-8}$ 
and are categorized as ``medium" decays. 
In general, the IB contribution dominates the decays, 
but the relative size is suppressed
if the intrinsic decay, with which the photon is associated, 
is suppressed or forbidden
(e.g. the $K^0_L\to\pi^+\pi^-$ decay by CP symmetry and 
the $K^+\to\pi^+\pi^0$ decay by the $\Delta I = 1/2$ rule).
A better understanding of radiative decays,
not only the branching ratios but also the decay spectra,
is important (and occasionally crucial)
to do rare decay search and precise measurement.
The $K^0_L \to \pi^0 \gamma \gamma$ decay
to the $K^0_L \to \pi^0 \ell \bar{\ell}$ predictions 
(Section~\ref{subsec:Rare-Kpll})
and
the $K^{\pm} \to e^{\pm} \nu\gamma$ decay
to the $K^{\pm}\to e^{\pm}\nu$ analysis
(Section~\ref{subsec:Lepton-LFU})
are good examples.

The IB component of a decay is a QED radiative correction to the intrinsic decay
and can be predicted reliably.
On the other hand,
the SD or DE component is
essentially 
due to low-energy hadronic interactions,
to which 
the effective-field approach based on chiral symmetry, 
{chiral perturbation theory}~\footnote{
   The techniques using chiral symmetry, called {\em chiral dynamics}, 
   were developed in the late 1960's
   and yield the results to meson interactions derived earlier 
   by {\em PCAC} and {\em current algebra}. 
   However, there had been no framework
   to do systematic calculations of hadron decays in the energy region below 1 GeV.
   In 1979, Weinberg introduced the concept of effective field theories
   using effective Lagrangians
   and proposed the application to strong interactions~\cite{Weinberg1,Weinberg2}.
   Gasser and Leutwyler, in 1984 and 1985, materialized Weinberg's program
   and expanded the use of chiral Lagrangian to the higher order~\cite{GL1,GL2}.
   This approach is called {\em chiral perturbation theory}, and 
   has been intensively used in particle and nuclear physics as a theory of low energy QCD.
   }~\cite{ChPTtext}, 
is applicable. 
In the theory, 
the quark fields are represented instead by the pseudoscalar-meson fields
and the interactions are constructed based on chiral symmetry
to the Nambu-Goldstone bosons.
The Lagrangian $\mathcal{L}$
is given in an expansion
in powers of derivatives ($\partial_{\mu}$) to the fields, 
i.e. their external momenta ($p$),
and with phenomenological coupling constants, and
the interactions vanish in the limit $p\to 0$.
By replacing the derivative in $\mathcal{L}$
to the covariant derivative ($D_{\mu}$),
the electroweak interactions are introduced to the theory and 
decay diagrams involving kaons and pions can be calculated.
Many applications to the medium kaon decays are collected
in the \DAFNE Physics Handbook~\cite{DAFNE1}. 
Since chiral perturbation theory 
is an effective field theory and does not make  ``perturbative" calculations
(and convergence) in the usual sense of Quantum Field Theory,
the theory should be verified by comparing the predictions with experiments,
including the measurements of radiative kaon decays.

\begin{table}
\begin{center}
\begin{minipage}[t]{16.5 cm}
\caption{
Radiative kaon decays in the last five years~\cite{PDG2010update}.}
\label{tab:radiativedecays}
\end{minipage}
  \begin{tabular}{l|cl|l|lll}
 \hline
 mode                                                  
                                                                  &  \multicolumn{2}{|l|}{branching ratio} 
                                                                        &  kinematic region$^{\ast}$   
                                                                                   & \multicolumn{2}{|l}{new results from}\\
\hline

$K^0_L\to \pi^{\pm} e^{\mp} \nu_{e} \gamma$  
                                                                 &  $( 3.79  \pm  0.06 )$   & $\times 10^{-3}$ 
                                                                        & $E_{\gamma}>30{\rm MeV}$, $\theta_{e\gamma}>20^{\circ}$
                                                                                  & KLOE   & \cite{KLOE-klpeng}\\
$K^0_L\to \pi^{\pm} e^{\mp} \nu e^+ e^-$  
                                                                 &  $( 1.26  \pm  0.04 )$   & $\times 10^{-5}$ 
                                                                        &  $M_{e e} > 5 {\rm MeV}$, $E_{e e} > 30 {\rm MeV}$
                                                                                  & KTeV   & \cite{KTeV-klpenee}\\
$K^0_L\to \pi^{0} \gamma \gamma$  
                                                                 &  $( 1.273  \pm  0.034 )$   & $\times 10^{-6}$ 
                                                                        &  
                                                                                  & KTeV   & \cite{KLggKTeVfinal}\\
$K^0_L\to \pi^{0} \gamma e^+ e^-$  
                                                                 &  $( 1.62  \pm  0.17 )$   & $\times 10^{-8}$ 
                                                                        &  
                                                                                  & KTeV   & \cite{KTeV-klpgee}\\
$K^0_L\to e^+ e^- \gamma$  
                                                                 &  $( 9.4  \pm  0.4 )$   & $\times 10^{-6}$ 
                                                                        &  
                                                                                  & KTeV   & \cite{KTeV-kleeg}\\

\hline

$K^0_S\to \pi^+ \pi^- e^+ e^-$  
                                                                 &  $( 4.79  \pm  0.15 )$   & $\times 10^{-5}$ 
                                                                        & 
                                                                                  & NA48/1   & \cite{NA48-ksppee}\\
$K^0_S\to \gamma\gamma$  
                                                                 &  $( 2.63  \pm  0.17 )$   & $\times 10^{-6}$ 
                                                                        & 
                                                                                  & KLOE   & \cite{KLOE-ksgg}\\

\hline

$K^+\to e^+ \nu_{e}\gamma$           
                                                                 & $( 9.4\   \pm  0.4\   )$   & $\times 10^{-6}$    
                                                                        &  $E_{\gamma}>10{\rm MeV}$, $p_{e}>200{\rm MeV}/c$ 
                                                                                  &  KLOE & \cite{KLOE-RK}  \\
$K^+\to \pi^0 e^+ \nu_{e} \gamma$    
                                                                 & $( 2.56  \pm  0.16 )$   & $\times 10^{-4}$ 
                                                                        & $E_{\gamma}>10{\rm MeV}$, $0.6< \cos(\theta_{e\gamma})<0.9$ 
                                                                                 & ISTRA+   & \cite{ISTRA-peng}  \\                                                                                                                              
$K^+\to \pi^0 \mu^+ \nu_{\mu} \gamma$    
                                                                 & $( 1.25  \pm  0.25 )$   & $\times 10^{-5}$ 
                                                                        & $30 < E_{\gamma} < 60 {\rm MeV}$
                                                                                 & E787   & \cite{E787-pmng}  \\
                                                                           
                                                                 &                                      &
                                                                        & 
                                                                                 & ISTRA+  & \cite{ISTRA-pmng}  \\
$K^+\to \pi^+ \pi^0 \gamma$\ \ (INT)    
                                                                 & $(-4.2  \pm  0.9 )$   & $\times 10^{-6}$ 
                                                                        & $0 < T_{\pi^+} < 80 {\rm MeV}$
                                                                                 & NA48/2   & \cite{NA48-ppg}\\
$K^+\to \pi^+ \pi^0 \gamma$\ \ (DE)    
                                                                 & $(6.0\  \pm  0.4\ )$   & $\times 10^{-6}$ 
                                                                        & $0 < T_{\pi^+} < 80 {\rm MeV}$
                                                                                 & NA48/2   & \cite{NA48-ppg}\\
$K^+\to \pi^+ e^+ e^- \gamma$  
                                                                 &  $( 1.19  \pm  0.13 )$   & $\times 10^{-8}$ 
                                                                        & $m_{e e \gamma} > 260 {\rm MeV}$
                                                                                 & NA48/2   & \cite{NA48-peeg}\\

 \hline
\end{tabular}
\begin{minipage}[t]{16.5 cm}
\vskip 0.5cm
\noindent
$^{\ast}$ Kinematic regions are defined in the kaon rest frame.
\end{minipage}
\end{center}
\end{table}

In table~\ref{tab:radiativedecays}, 
radiative neutral- and charged-kaon decays
whose new branching-ratio measurement was published 
in the last five years are listed.
Some of them were measured as a byproduct of the experiment
with the data collected by a pre-scaled physics or calibration trigger. 
It should be noted that,
since the detector system has an energy threshold for photon detection
and, also in the theoretical calculation,  
a cutoff to the photon energy is set 
in order to avoid infrared divergence to the IB component,
a radiative decay is usually measured 
in a specific kinematic region.

\begin{figure}[tb]
\begin{center}
\begin{minipage}[t]{16.5cm}
\begin{center}
\epsfig{file=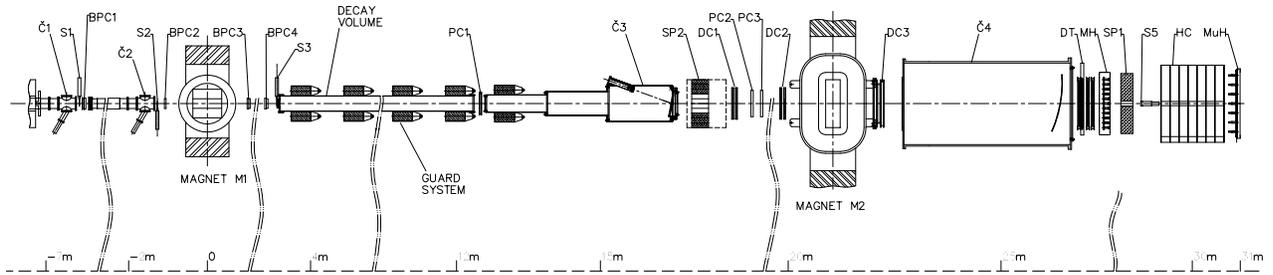,scale=0.60,angle=90}
\end{center}
\end{minipage}
\begin{minipage}[t]{16.5cm}
\caption{
Elevation view of the ISTRA+ detector.
BPC1-BPC4: multiwire chambers for tracking beam particles;
\v{C}1 and \v{C}2: threshold \v{C}erenkov counters  for kaon identification;
S1-S3: scintillating counters for beam trigger, 
SP2: lead-glass calorimeter to detect large-angle photons,
PC1-PC3, DC1-DC3, and DT: proportional chambers, drift chambers and drift tubes, respectively,
for tracking decay products;
\v{C}3 and \v{C}4: wide-aperture threshold \v{C}erenkov counters;
MH: scintillating hodoscope;
SP1, HC, and MuH: lead-glass calorimeter, 
iron-scintillator sampling hadron calorimeter, and 
muon hodoscope, respectively, for decay products; 
S5: scintillating counter at the downstream of the setup to veto beam particles which do not decay.
Source: Figure taken from Ref.\cite{ISTRA-mng}.
\label{Fig:ISTRAdet}
}
\end{minipage}
\end{center}
\end{figure} 

The ISTRA+ collaboration at IHEP
 measured $K^-$ decays in flight
 with the 26 GeV/$c$ un-separated secondary beam
 from U-70, 
 the IHEP 70GeV proton synchrotron, in 2001. 
 The ISTRA+ detector,
 with the spectrometer, 
 lead-glass electromagnetic calorimeter (SP$_1$) and sampling hadron calorimeter (HC),
  is shown in Fig.~\ref{Fig:ISTRAdet}. 
  They reported the measurements of 
  $K^-\to \pi^0 e^- \nu_{e} \gamma$~\cite{ISTRA-peng} and
  $K^-\to \pi^0 \mu^- \nu_{\mu} \gamma$~\cite{ISTRA-pmng}.
 
As a successor to  ISTRA+, 
a new ``OKA" experimental program~\cite{OKA} 
with RF-separated \Kpm beam
is ongoing at U-70 of IHEP.

\clearpage

\section{Hadrons in kaon decays
\label{sec:hadron}}

As a theory of low energy QCD, 
the most reliable predictions of chiral perturbation theory
are the S-wave $\pi\pi$ scattering lengths 
$a_I$ in the isospin $I=0$ and $I=2$ states~\cite{CGL2000,CGL2001}: 
$a_0 = 0.220 \pm 0.005$,  
$a_2 = -0.0444 \pm 0.0010$, and
$(a_0 - a_2) = 0.264 \pm 0.004$ 
in units of $1/m_{\pi^+}$, 
where $m_{\pi^+}$ is the charged-pion mass. 
The DIRAC collaboration at CERN produced $\pi^+\pi^-$ atoms
to measure its lifetime, and 
recently obtained
$|a_0-a_2| = 0.2533\ ^{+0.0080}_{-0.0078}(stat.)\ ^{+0.0078}_{-0.0073}(syst.)$~\cite{DIRAC2011}.
The $\pi\pi$ scattering lengths can also be determined
by precise measurement of the kaon decays
with two pions in the final state.

\begin{figure}[tb]
\begin{center}
\begin{minipage}[t]{16.5 cm}
\begin{center}
\epsfig{file=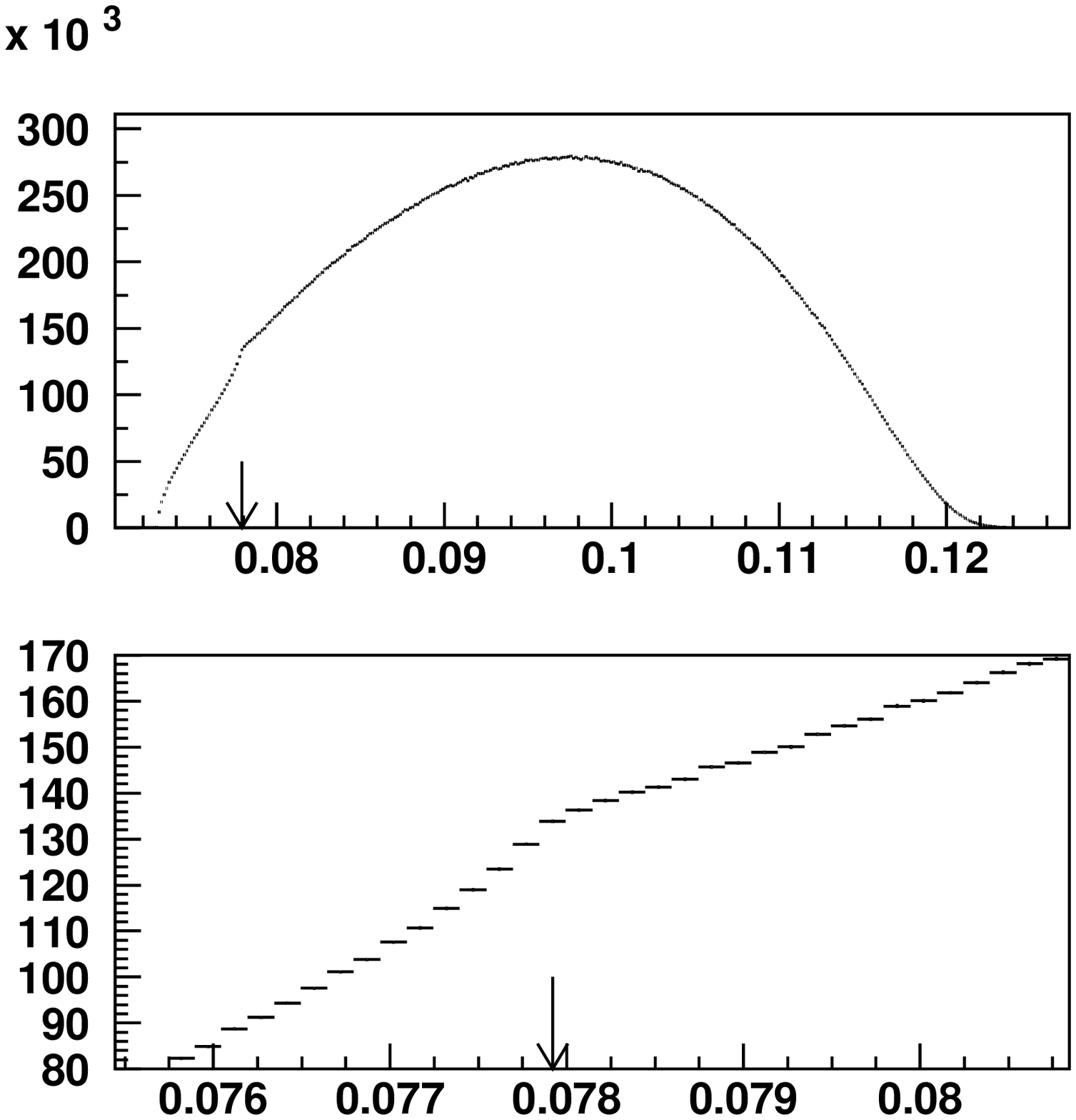,scale=0.40}
\hspace*{0.5cm}
\epsfig{file=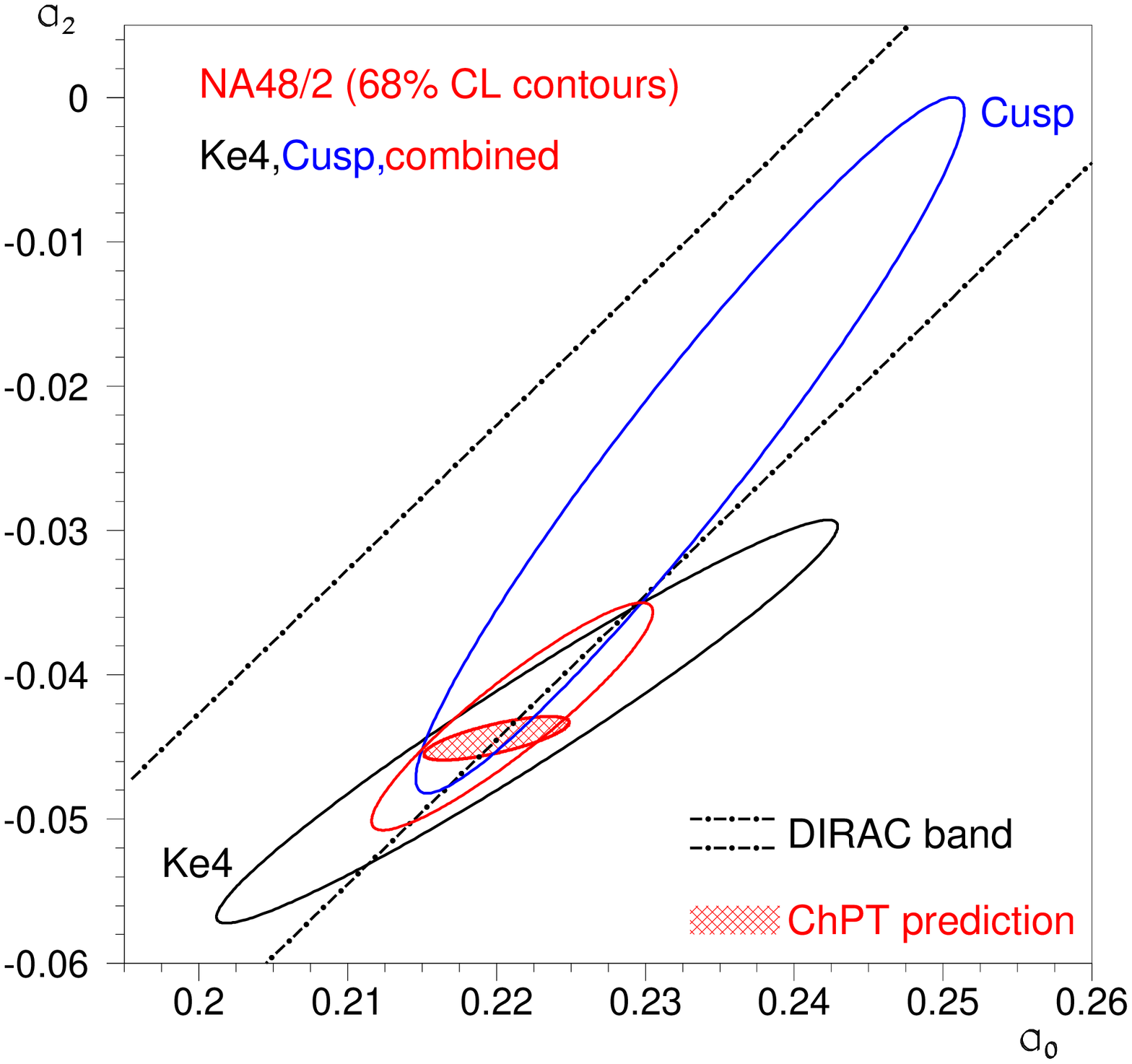,scale=0.35}
\end{center}
\end{minipage}
\begin{minipage}[t]{16.5cm}
\caption{
Cusp in $K^{\pm}\to \pi^{\pm}\pi^+\pi^-$ by NA48/2: 
distribution of the square of the $\pi^0\pi^0$ invariant mass, $M_{00}^2$, 
in units of (GeV/$c^2$)$^2$ and
enlargement of a narrow region 
around $M_{00}^2= (2m_{\pi^+})^2$ (left); 
NA48/2 $K_{e 4}$ and cusp results from two-parameter fits
in the ($a_0$,$a_2$) plane (right).
In the right figure, 
the smallest contour corresponds to
the combination of NA48/2 results,
the cross-hatched ellipse is the prediction of
chiral perturbation theory,
and the dashed-dotted lines correspond to 
the recent results from the DIRAC experiment~\cite{DIRAC2011}. 
Source: Figures taken from Ref.\cite{NA48cuspfinal} (left)
and 
by courtesy of B. Bloch-Devaux (right).
\label{Fig:cuspNA48}
}
\end{minipage}
\end{center}
\end{figure}

In the $\pi^0\pi^0$ invariant-mass ($M_{00}$) distribution
in the $K^{\pm}\to \pi^{\pm}\pi^0\pi^0$ decay,
the NA48/2 experiment observed
a {\em cusp}-like anomaly  in the region near $M_{00}= 2m_{\pi^+}$
(Fig.~\ref{Fig:cuspNA48} left).
The cusp was interpreted as an effect
due to the interference of 
the two $K^{\pm}\to \pi^{\pm}\pi^0\pi^0$ amplitudes; 
the first one is the intrinsic $K^{\pm}\to \pi^{\pm}\pi^0\pi^0$ transition
and the second one is $K^{\pm}\to \pi^{\pm}\pi^+\pi^-$
followed by final state rescattering $\pi^+\pi^-\to \pi^0\pi^0$~\cite{Cabibbo04,CI05}.
Fits to the $M_{00}$ distribution from the 2003 and 2004 data sets 
using two formulations 
with several theoretical issues~\cite{CGKR06,GTV07,BFGKR2009}
taken into account, 
NA48/2 obtained
the results from the cusp of $K^{\pm}\to \pi^{\pm}\pi^0\pi^0$ as
$(a_0-a_2) = 0.2571 \ \pm 0.0048(stat.)\ \pm 0.0025(syst.)\ \pm 0.0014(ext.)$
and
$a_2 = -0.024 \ \pm 0.013(stat.)\ \pm 0.009(syst.)\ \pm 0.002(ext.)$~\cite{NA48cuspfinal}.

The properties of the $K^{\pm}\to \pi^+ \pi^- e^{\pm} \nu$ ($K_{e 4}$) decay, 
which has no other hadron than two pions in the final state, 
were measured by the NA48/2 experiment
based on the statistics, 1.13M events from the 2003 and 2004 data sets,
better than the previous experiments~\cite{E865Ke4,S118Ke4}. 
As a result of the analysis, 
the difference (shift) $\delta$
between 
the S-wave, $I=0$ and 
P-wave, $I=1$ form-factor phases~\cite{PaisTreiman}, 
$\delta^{0}_{0}$ and $\delta^{1}_{1}$, 
was obtained 
as a function of the $\pi^+\pi^-$ invariant mass.
The phase shift $ \delta = ( \delta^{0}_{0} - \delta^{1}_{1} ) $
is related to the $\pi\pi$ scattering lengths
by using numerical solutions~\cite{ACGL01,DFGS02}
of the Roy equations~\cite{Roy71},
which are a set of dispersion relations for the partial wave amplitudes of 
$\pi\pi$ scattering.
After taking the isospin-breaking effects~\cite{CGR2009} into account, 
NA48/2 obtained
$a_0  = 0.2220 \ \pm 0.0128(stat.)\ \pm 0.0050(syst.)\ \pm 0.0037(th.)$
and
$a_2 = -0.0432 \ \pm 0.0086(stat.)\ \pm 0.0034(syst.)\ \pm 0.0028(th.)$~\cite{NA48Ke4results}
from the $K_{e 4}$ decay.
Finally,
combining both the $K_{e 4}$ and cusp results from NA48/2 
(Fig.~\ref{Fig:cuspNA48} right)
without using the relation between $a_0$ and $a_2$ from chiral perturbation theory, 
$a_0  = 0.2210 \ \pm 0.0047(stat.)\ \pm 0.0040(syst.)$, 
$a_2 = -0.0429 \ \pm 0.0044(stat.)\ \pm 0.0028(syst.)$, and
$(a_0-a_2) = 0.2639 \ \pm 0.0020(stat.)\ \pm 0.0015(syst.)$
were obtained~\cite{NA48Ke4results}
as the most precise tests of low energy QCD from the kaon decay properties.

The Dalitz plot density of $K^{\pm}\to \pi^{\pm}\pi^0\pi^0$
should be described 
with the known cusp in the region near $M_{00}= 2m_{\pi^+}$;
the NA48/2 experiment proposed a new empirical model-independent parametrization 
of the $K^{\pm}\to \pi^{\pm}\pi^0\pi^0$ Dalitz plot
with seven parameters, and provided the values obtained by fitting their data~\cite{NA48-empparam}.
Rescattering effects are much smaller
in the  $K^{\pm}\to \pi^{\pm}\pi^+\pi^-$ decay~\footnote{ 
     Since the invariant mass of any two pion pair in $K^{\pm}\to \pi^{\pm}\pi^+\pi^-$ is always
     $\geq 2m_{\pi^+}$, any cusp structure in the decay is outside the physical region.}; 
the parametrization in Eq.~(\ref{eq:kp3sq}) is still valid, 
and 
the NA48/2 experiment
measured the values of the slope parameters~\cite{NA48-slopeparam}.
NA48/2 also obtained, 
concurrently with their phase-shift measurements, 
the form factors 
in the matrix element of the $K_{e 4}$ decay~\cite{NA48Ke4results,BBD2011}.

\begin{figure}[tb]
\begin{center}
\begin{minipage}[t]{16.5 cm}
\begin{center}
\epsfig{file=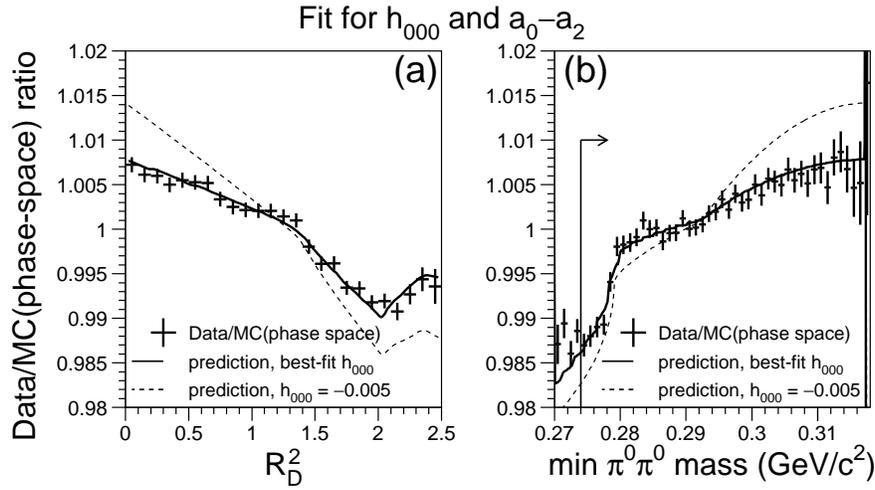,scale=0.60}
\end{center}
\end{minipage}
\begin{minipage}[t]{16.5cm}
\caption{
Cusp in $K^0_L \to \pi^0\pi^0\pi^0$ by KTeV: 
distributions of the data/MC(phase-space) ratio 
as a function of $R_D^2$ (left)
and the minimum $\pi^0\pi^0$ mass (right).
The solid curve is the prediction from the best fit, and 
the dashed curve is the prediction from the Dalitz-plot density with   $h_{000} = -0.005$ 
without the rescattering effect.
Source: Figures taken from Ref.\cite{KTeVcuspfinal}.
\label{Fig:cuspKTeV}
}
\end{minipage}
\end{center}
\end{figure}

In the $K^0_L \to \pi^0\pi^0\pi^0$ decay,
the KTeV experiment also observed
a cusp in the region 
where the the minimum $\pi^0\pi^0$ invariant-mass is near $2m_{\pi^+}$, 
though the cusp is smaller than in $K^{\pm}\to \pi^{\pm}\pi^+\pi^-$
and is visible  (Fig.~\ref{Fig:cuspKTeV})
in the distribution of the ratio to the phase-space Monte Carlo (MC).
The $K^0_L \to \pi^0\pi^0\pi^0$ Dalitz-plot density is 
parametrized as
 \be
 | M(R_D) |^2 \ \propto\  1\ +\ h_{000}\ \cdot\ R_{D}^{2}\ , \label{eq:kp3000sq}
\ee 
where the coefficients $h_{000}$ is the quadratic slope parameter
and
$R_{D}$
is the distance from the center of the Dalitz plot.
Using the model in \cite{CI05} and
fitting the data for both $h_{000}$ and $(a_0-a_2)$
in a two-parameter fit,
KTeV obtained
$h_{000} = (-2.09\ \pm 0.62(stat.)\ \pm 0.72(syst.)\ \pm 0.28(ext.))\times 10^{-3}$
and 
$(a_0-a_2) = 0.215 \ \pm 0.014(stat.)\ \pm 0.025(syst.)\ \pm 0.006(ext.)$~\cite{KTeVcuspfinal}
from $K^0_L \to \pi^0\pi^0\pi^0$.

New form factor measurements were reported
by KLOE~\cite{KLOE-pen} 
on $K^0_L\to \pi^{\pm}e^{\mp}\nu$
and by KTeV~\cite{KTeV-disp}, KLOE~\cite{KLOE-pmn} and NA48~\cite{NA48-pmn}
on $K^0_L\to \pi^{\pm}\mu^{\mp}\nu$
as well as
by KLOE~\cite{KLOE-RK}
on $K^{\pm} \to e^{\pm} \nu \gamma$
and by ISTRA+~\cite{ISTRA-mng}
on $K^- \to \mu^- \nu \gamma$.

\clearpage

\section{Basic observables
\label{sec:Basic}}

\begin{table}
\begin{center}
\begin{minipage}[t]{16.5 cm}
\caption{
Kaon basic observables$^{\ast}$.}
\label{tab:basicobs}
\end{minipage}
  \begin{tabular}{ll|l|l|l}
 \hline
observable                & unit   
                                            & value in PDG2010~\cite{PDG2010update}
                                            & value in PDG2000~\cite{PDG2000}
                                                                                              &  new results from \\
\hline

$K^0$ mass: $m_{K^0}$                  &    MeV/$c^2$    
                                             & $497.614 \pm 0.024$
                                             & $497.672 \pm 0.031$
                                             & KLOE, NA48 \\
\hline
$K^0_L$ mean life: $\tau_{K^0_L}$       &   nsec    
                                             & $51.16 \pm 0.21 $
                                             & $51.7   \pm 0.41$
                                             & KLOE \\
$\ K^0_L\to \pi^{\pm}e^{\mp}\nu_{e}$  & $10^{-2}$
                                             & $40.55 \pm 0.11$
                                             & $38.78 \pm 0.28$
                                             & KLOE, KTeV \\
$\ K^0_L\to \pi^{\pm}\mu^{\mp}\nu_{\mu}$ & $10^{-2}$
                                              & $27.04 \pm 0.07$
                                              & $27.18 \pm 0.25$
                                              & KLOE, KTeV \\
$\ K^0_L\to \pi^0\pi^0\pi^0$  &  $10^{-2}$
                                              & $19.52 \pm 0.12$ 
                                              & $21.13 \pm 0.27$
                                              & KLOE, KTeV \\
$\ K^0_L\to \pi^+\pi^-\pi^0$  &  $10^{-2}$
                                              & $12.54 \pm 0.05$
                                              & $12.55 \pm 0.20$
                                              & KLOE, KTeV \\
$\ K^0_L\to \pi^+\pi^-$  &  $10^{-3}$
                                               & $1.966 \pm 0.010$
                                               & $2.056 \pm 0.033$
                                               & KTeV \\
$\ K^0_L\to \pi^0\pi^0$  & $10^{-4}$
                                               & $8.65 \pm 0.06$
                                               & $9.27 \pm 0.19$
                                               & KTeV \\
\hline
$K^0_S$ mean life: $\tau_{K^0_S}$       &   psec    
                                             & $89.53 \pm 0.05 $
                                             & $89.35  \pm 0.08$
                                             & KTeV, NA48 \\
$\ K^0_S\to \pi^0\pi^0$  & $10^{-2}$
                                               & $30.69 \pm 0.05$
                                               & $31.39 \pm 0.28$
                                               & KLOE \\
$\ K^0_S\to \pi^+\pi^-$  & $10^{-2}$
                                               & $69.20 \pm 0.05$
                                               & $68.61 \pm 0.28$
                                               & KLOE \\                                             
$\ K^0_S\to \pi^{\pm}e^{\mp}\nu_{e}$  & $10^{-4}$
                                             & $7.04 \pm 0.08$
                                             & $7.2 \pm 1.4$
                                             & NA48, KLOE \\
\hline
$|\epsilon|$                        & $10^{-3}$
                                                 & $2.228 \pm 0.011$
                                                 & $2.271 \pm 0.017$
                                                 & (from fit)\\                                                                                                                                               
\hline
$K^+$ mass: $m_{K^+}$                  &    MeV/$c^2$    
                                             & $493.677 \pm 0.016$
                                             & $493.677 \pm 0.016$
                                             & \\
$K^+$ mean life: $\tau_{K^+}$            &   nsec    
                                             & $12.380 \pm 0.021 $
                                             & $12.386  \pm 0.024$
                                             & KLOE \\
$\ K^+\to \mu^{+} \nu_{\mu}$  & $10^{-2}$
                                             & $63.55 \pm 0.11$
                                             & $63.51 \pm 0.18$
                                             & KLOE \\
$\ K^+\to \pi^0 e^+ \nu_{e}$    &  $10^{-2}$
                                             & $5.07 \pm 0.04$
                                             & $4.28 \pm 0.06$
                                             & KLOE \\
$\ K^+\to \pi^{0}\mu^+ \nu_{\mu}$  &  $10^{-2}$
                                             & $3.353 \pm 0.034$
                                             & $3.18 \pm 0.08$
                                             & KLOE \\
$\ K^+\to \pi^+\pi^0$        & $10^{-2}$
                                             & $20.66 \pm 0.08$
                                             & $21.16 \pm 0.14$
                                             & KLOE \\
$\ K^+\to \pi^+\pi^0\pi^0$  & $10^{-2}$
                                              & $1.761 \pm 0.022$
                                              & $1.73 \pm 0.04$
                                              & KLOE \\
$\ K^+\to \pi^+ \pi^+ \pi^-$  &  $10^{-2}$
                                              & $5.59 \pm 0.04$
                                              & $5.59 \pm 0.05$
                                              & \\
\hline
$|V_{us}|$                        & 
                                                 & $0.2252 \pm 0.0009$
                                                 & $0.2196 \pm 0.0023$
                                                 & Section~\ref{sec:Vus} \\
 \hline
\end{tabular}
\begin{minipage}[t]{16.5 cm}
\vskip 0.5cm
\noindent
$^{\ast}$ Decay modes mean their branching ratios.
\end{minipage}
\end{center}
\end{table}

Kaon decay experiments remeasured 
basic observables such as kaon masses, lifetimes, and absolute branching ratios.
Features of the modern kaon experiments,
with mature experimental methods, 
are
the large statistics (high intensity of the beam and large acceptance of the detector),
high resolution (state-of-the art and sophisticated techniques in the detector),
advanced computing (for both data analysis and MC simulation),
controlled systematic-uncertainties (e.g. simultaneous measurement of multiple decay modes)
and better handling of the theoretical issues 
such as radiative corrections~\cite{Gatti06,BI06radcorr,Andre07,NW2007PHOTOS,Cirigliano06}.
All of these are helpful in understanding the detector, background sources 
and uncertainties
to do precise measurement of ``(should-have-been) well-done" decays. 

Instead of discussing each measurement,
the values of basic observables taken from the books of {Review of Particle Physics}
in 2010~\cite{PDG2010update} and in 2000~\cite{PDG2000} are listed
in table~\ref{tab:basicobs},
to present how precisions were improved
(or central values were significantly shifted)
in the last ten years.
It should also be mentioned that
new measurements have come 
after {Review of Particle Physics} in 2010,
for example
on the $K^0_S$ mean life as
$ ( 89.562 \pm 0.029(stat.) \pm 0.043(syst.) ) $ psec from KLOE~\cite{KLOE-KSlife}
and
$ ( 89.623 \pm 0.047 ) $ psec from KTeV~\cite{KTeV-final}.

It has been pointed out
(e.g. \cite{LunghiSoni08,BurasGuadagnoli08})
that an inconsistency exists in $|\epsilon|$
between the experimental value
($ ( 2.228 \pm 0.011 ) \times 10^{-3}$ in \cite{PDG2010update})
and the SM prediction
($ ( 1.81 \pm 0.28 ) \times 10^{-3}$ in \cite{1108.2036}).
This can be regarded as a tension 
between $|\epsilon|$ and 
$\sin 2\phi_1$ in the CKM-parameter fits,
indicating hints for New Physics effects.

\clearpage

\section{$V_{us}$ and CKM unitarity
\label{sec:Vus}}

All the roads of kaon physics lead to CKM, 
in particular the $V_{us}$ fountain,
which was not built in a day. 
The magnitude of $V_{us}$, 
or $\lambda$ in the Wolfenstein parametrization
or  $\sin \theta_C$ with the Cabibbo angle $\theta_C$, 
is extracted from semileptonic kaon decays
with
\bea
  B( K_{\ell 3} ) & = &
    \tau_K\ \times\ \frac{ G_{F}^2\ m_{K}^5 }{ 192\ \pi^3}\ C_{K}^2\ S_{EW}\  
     \times\ |V_{us}|^2 \cdot  f_+(0)^2\ \times\ I_{K\ell}\ \left(\ 1\ +\ \delta_{EM}^{K\ell}\ +\ \delta_{SU(2)}^{K\pi}\ \right)^2 
\label{eq:Kl3CKM}\\
                           & \propto &
     \left[\ \ |V_{us}| \cdot f_+(0)\ \ \right]^2\ 
\eea
or leptonic kaon decays
with
\bea
  \frac{ B(K_{\ell 2})  }{ B(\pi_{\ell 2})  } & = &
    \frac{ \tau_K }{ \tau_{\pi} }\ \times\  \frac{ |V_{us}|^2 }{ |V_{ud}|^2 } \cdot \frac{ f_{K}^2 }{ f_{\pi}^2 }\ \times\ 
    \frac{ m_K\ (\ 1\ -\  m_{\ell}^2 / m_{K}^2\ )^2 }{ m_\pi\ (\ 1\ -\  m_{\ell}^2 / m_{\pi}^2\ )^2 }\ 
    \left(\ 1\ +\ \delta_{EM}\ \right) 
\label{eq:Kl2CKM}\\
                                                                   & \propto &
    \left[\ \ \frac{ |V_{us}| }{ |V_{ud}| } \cdot \frac{ f_{K} }{ f_{\pi} }\ \ \right]^2\ .
\eea
In Eq.~\ref{eq:Kl3CKM}, 
$G_F$ is the Fermi constant, $C_K^2$ is $1$ for $K^0$ decay and $1/2$ for $K^{\pm}$ decay, 
$S_{EW}=1.0232(2)$ is the short-distance electroweak correction, 
$I_{K \ell}$ is a phase-space integral that is sensitive to the momentum dependence of the form factors, 
$\delta_{EM}^{K\ell}$ is the channel-dependent long-distance electromagnetic correction and
$\delta_{SU(2)}^{K\pi}$ is the correction for isospin breaking 
(and is zero for $K^0$ decay)~\footnote{\ 
$(1\ +\ \delta_{EM}^{K\ell}\ +\ \delta_{SU(2)}^{K\pi})^2\ \simeq\ 
  (1\ +\ 2\ \delta_{EM}^{K\ell}\ +\ 2\ \delta_{SU(2)}^{K\pi})$\ 
is used in some papers.
}.
In Eq.~\ref{eq:Kl2CKM}, 
$\delta_{EM}= -0.0070(18)$
is the long-distance electromagnetic correction
that does not cancel from the ratio. 
What we obtain
from the semileptonic and leptonic kaon-decay measurements
are the values of $ |V_{us}| \cdot f_+(0)$,
where $f_+(0)$ is the $K\to \pi$ vector form factor at zero momentum transfer,
and
the product of the ratios $ |V_{us}| /  |V_{ud}| $ and $ { f_{K} }/{ f_{\pi} }$,
where
$  f_{K} $ and $ f_{\pi} $ are the kaon and pion decay constants, respectively.
The values of $f_+(0)$ and $ { f_{K} }/{ f_{\pi} }$ should be provided
in order to extract $|V_{us}|$; 
they are obtained theoretically 
from lattice QCD calculations. 

The $|V_{us}|$ results should satisfy 
one of the unitarity conditions of the CKM matrix elements, 
$ |V_{ud}|^2 + |V_{us}|^2 + |V_{ub}|^2 = 1$.
On the other hand, 
a new parameter $\Delta_{CKM}$ defined as
\be
   \Delta_{CKM} \equiv  |V_{ud}|^2\ +\ |V_{us}|^2\ +\ |V_{ub}|^2\ -\ 1\ \label{eq:DeltaCKM}
\ee
is a benchmark to test the consistency of quark mixing in the SM. 
It had been pointed out that 
$|V_{us}| \simeq 0.220$
(the value in PDG2000, in table~\ref{tab:basicobs})
leaded to $\Delta_{CKM} = (-3.2 \pm 1.4)\times 10^{-3}$ 
and indicated a two-sigma deviation from zero, 
until the BNL E865 experiment, 
whose main purpose was a search for $K^+ \to \pi^+ \mu^+ e^-$~\cite{E865kpmn},
published a new measurement of 
$K^+\to \pi^0 e^+ \nu$~\cite{E865pen}
and claimed the branching ratio (and $|V_{us}|$) was  larger.  
After that, more new results of 
precise branching-ratio measurements on neutral and charged kaon decays came (Section~\ref{sec:Basic}),
and there have been laborious works in the collaboration of theorists and experimentalists,
in particular by the FlaviaNet Kaon Working Group~\cite{FlaviaNetK}, 
to evaluate $|V_{us}| $ precisely.
The latest achievements are described in details in \cite{EPJC69-399},
and a compact review is available in \cite{PDG-Vus}.
Instead of repeating these contents, 
the final values in  \cite{EPJC69-399} are summarized.
The experimental result on $ |V_{us}| \cdot f_+(0)$ is $0.2163(5)$ and, 
with $f_+(0) = 0.959(5)$, 
$ |V_{us}|$ is $0.2254(13)$  from semileptonic kaon decays.
The experimental result on  
$ |V_{us}| /  |V_{ud}| \cdot { f_{K} }/{ f_{\pi} }$ is $0.2758(5)$ and,
with ${f_{K} }/{ f_{\pi} } =1.193(6)$, 
$ |V_{us}| /  |V_{ud}| $ is $0.2312(13)$ from leptonic kaon decays. 
By using $ |V_{ud}| = 0.97425(22)$, 
the combined $ |V_{us}|$ is $0.2253(9)$ 
and  $\Delta_{CKM}$ is $-0.0001(6)$.

The efforts on $ |V_{ud}| $ will continue 
with the results from new kaon experiments.

\clearpage

\section{Exotic searches
\label{sec:Exotic}}

Experimental searches for very light bosons have a long history, but
a neutral boson whose mass is twice the muon mass has not yet been 
excluded.
In 2005, the HyperCP collaboration at FNAL reported
three events of the $\Sigma^+\to p \mu^+ \mu^-$ decay, 
and the dimuon mass may indicate a neutral intermediate state $P^0$ 
with a mass of $214.3\pm 0.5$ MeV/$c^2$~\cite{HyperCP}.
Since the events were observed in an FCNC process with a strange to down quark transition,
$P^0$ should be confirmable with kaon decays. 
Dimuon masses in previous $K^{+}\to\pi^{+}\mu^+\mu^-$  measurements
were not observed in the narrow range around the $P^0$ mass; 
thus, $P^0$ should be a pseudo-scalar or axial-vector particle 
and be studied with the three-body decay $K\to\pi\pi P^0$.
The KTeV experiment searched for the $K^0_L \to \pi^0\pi^0\mu^+\mu^-$ decay
for the first time~\cite{KTeV-ppmm} and set 
$B(K^0_L \to \pi^0\pi^0\mu^+\mu^-) < 9.2\times 10^{-11} $
and 
$B(K^0_L \to \pi^0\pi^0 P^0\to\pi^0\pi^0\mu^+\mu^-) < 1.0\times 10^{-10}$.
The E391a experiment searched for 
a pseudoscalar particle $X^0$ 
in the decay $K^0_L \to \pi^0\pi^0 X^0$, $X^0\to\gamma\gamma$ 
in the mass range of $X^0$ from 194.3 to 219.3 MeV/$c^2$, 
and set 
$B(K^0_L \to \pi^0\pi^0 P^0\to\pi^0\pi^0\gamma\gamma) < 2.4\times 10^{-7}$~\cite{E391a-ppgg}.
Both results almost ruled out the predictions 
that
$P^0$ is a pseudo-scalar particle~\cite{HTV07,OT09}. 

The E787/E949 results on $K^+ \to \pi^+\nu\bar{\nu}$
have also been interpreted 
as a
two-body decay $K^+\to\pi^+ U^0$, 
where $U^0$ is a massive non-interacting particle 
either stable or unstable, 
and as $K^+ \to \pi^+ V^0$, $V^0\to\nu\bar{\nu}$,
for a hypothetical, short-lived particle $V^0$.
The limits are presented in~\cite{E949final}.
Limits on the three-body decays 
$K^-\to \pi^- \pi^0 S^0$ and $K^0_L\to \pi^0 \pi^0 S^0$ 
were obtained
in the ISTRA+ results~\cite{ISTRA-sgoldstino}
and
in the E391a results~\cite{E391appnn}, respectively,
as searches for 
light pseudoscalar sgoldstino-particles $S^0$~\cite{sgoldstino}.
The discovery potential of the rare FCNC kaon decays
with {\em missing energy} in the final state
is discussed recently in \cite{1111.6402}.

The NA48/2 experiment published the upper limit 
$B(K^{\pm} \to \pi^{\mp} \mu^{\pm} \mu^{\pm} ) < 1.1\times 10^{-9}$~\cite{NA48-pimm}.
This is a neutrino-less ``double muon" decay by changing total lepton number by two,
and provides a unique channel to search for effects of Majorana neutrinos
in the second generation of quarks and leptons~\cite{Zuber,LS2000}.

The ISTRA+ experiment performed a search for a heavy neutrino
in the mass from 40 to 80 MeV/$c^2$ and the lifetime from $10^{-11}$ to $10^{-9}$ 
in the $K^- \to \mu^- \nu_{h} \to \mu^- \nu \gamma$ decay,
and put upper limits on the square of the mixing matrix element $|U_{\mu h}|^2$~\cite{ISTRA-nu}.
Heavy neutrinos which  are a part of the $\nu_{\mu}$ flavor eigenstate and 
decay {\em radiatively} into a massless neutrino and a photon
are considered as an explanation to the long-standing LSND/KARMEN/MiniBooNE anomaly
in neutrino physics~\cite{Gninenko2009,Gninenko2011,Gninenko2011-2}.
This decay can be studied with the detector which measures the $K\to\mu\nu\gamma$ decay. 

The decay $K^0_L\to 3\gamma$, which is forbidden by charge-conjugation invariance
as in the case with $\pi^0\to 3\gamma$, can proceed via parity-violating interactions
without violating CP,
 but is further suppressed by the gauge invariance and Bose statistics~\cite{Yang1950}.
E391a published the new upper limit
$B(K^0_L \to 3\gamma) < 7.4\times 10^{-8}$~\cite{E391a-3g}
on the assumption that the decay proceeded via parity-violation.

The decay $K^+\to\pi^+\gamma$ is a spin $0\to 0$ transition with a real photon
and is thus forbidden by angular momentum conservation; 
it is also forbidden by gauge invariance.
But this decay is allowed in noncommutative theories~\cite{MPKT2005}. 
E949 published the upper limit 
$B(K^+ \to \pi^+ \gamma) < 2.3\times 10^{-9}$~\cite{E949-pgg}.

\clearpage

\section{Conclusion
\label{sec:Conclusion}}

The study of kaon physics continues to make great strides. 
The current program to study CP violation is being completed; 
the CP violation in \Repoe
was established, but 
the CP asymmetries in charged kaon decays have not been observed yet. 
The rare decays $K^0_L \to \pi^0\nu\bar{\nu}$ and $K^+ \to \pi^+\nu\bar{\nu}$,
the transverse muon polarization in $K^+$ decays at rest 
and the lepton flavor universality in 
$\Gamma(K^+\to\ e^+\nu(\gamma)) / \Gamma(K^+\to\ \mu^+\nu(\gamma))$
will be measured in a new series of experiments.
Tests of CPT and quantum mechanics will continue 
in $\phi$ factory experiments. 
New results on radiative kaon decays, $\pi\pi$ scattering lengths, 
Dalitz-plot densities, form factors, basic observables, $V_{us}$,
and exotic decays have been reported, 
and their measurements will continue in new experiments. 
The kaon experiments, with ultra-high sensitivities and precisions, 
will continue to be essential and crucial 
as a probe of New Physics beyond the Standard Model.

\section*{Acknowledgments}

I would like to thank
     B. Bloch-Devaux, 
     E. Blucher,
     F. Bossi, 
     A.J. Buras, 
     A. Ceccucci,
     A.~Di Domenico,
     V.A. Duk, 
     E. Goudzovski, 
     G. Isidori, 
     L.S. Littenberg,
     F. Mescia,
     M. Moulson, 
     C. Smith,
and
     T. Yamanaka
for providing me help with this article.
I would like to acknowledge support from 
Grant-in-Aids for Scientific Research on Priority Areas:
"New Developments of Flavor Physics" 
and for Scientific Research (S)
by the MEXT Ministry of Japan. 

 This article is dedicated to the memory of my wife, Yuko Nitta-Komatsubara,
 who passed away on March 21, 2011.

\end{document}